\documentclass[aip,apl,reprint]{revtex4-1}

\usepackage{graphicx}
\usepackage[alsoload=synchem]{siunitx}
\usepackage{braket}
\usepackage{xspace}
\usepackage{amsmath}
\usepackage{mathtools}

\DeclarePairedDelimiter{\abs}{\lvert}{\rvert}

\begin{document}

\def \Concept {
\begin{figure}[!t]
\includegraphics[width=\linewidth]{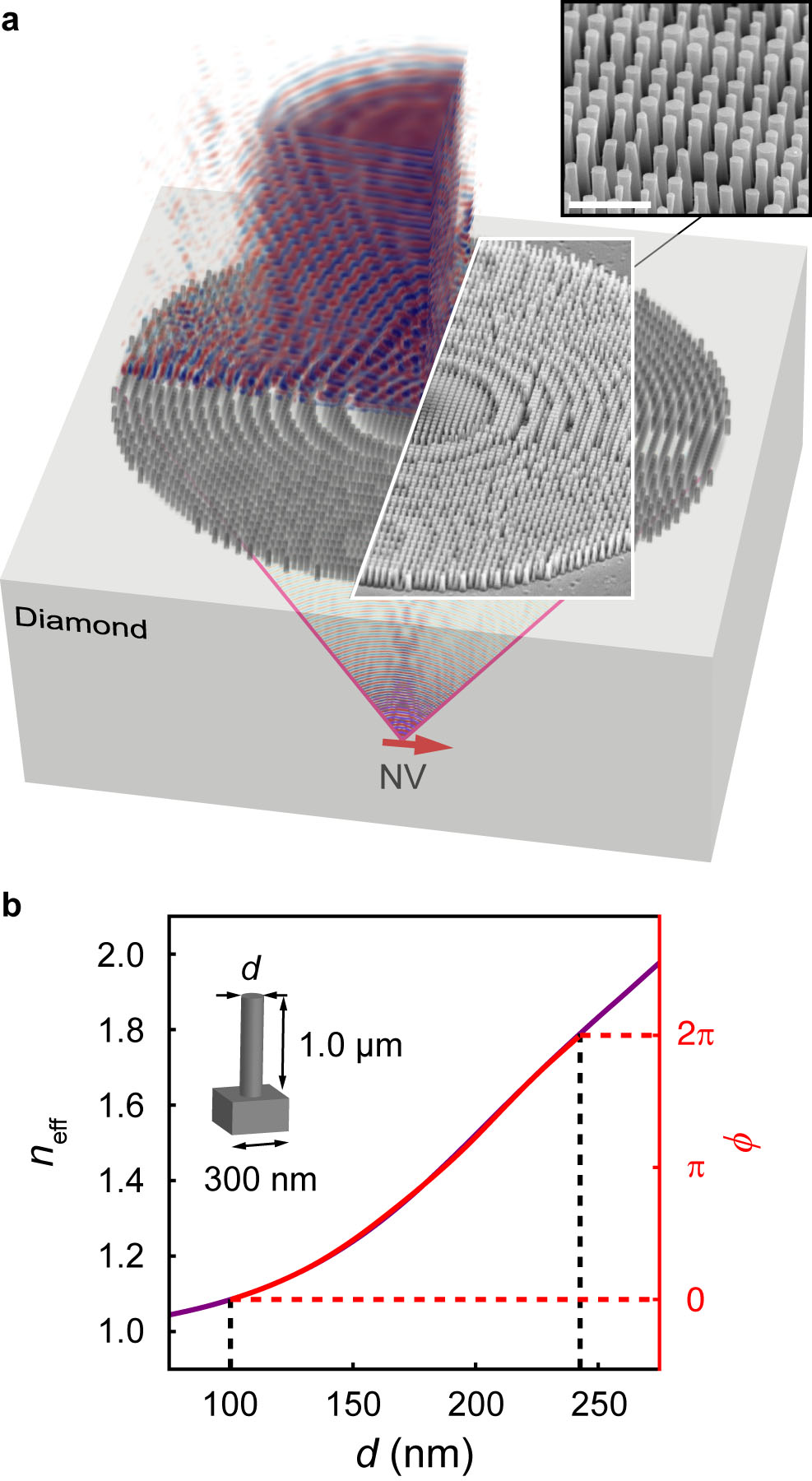}
\caption{\label{Fig:Concept}\textbf{Diamond immersion metalens.} \textbf{a}, Subwavelength pillars extending from the surface of a single-crystal diamond substrate are designed to create a high-numerical-aperture immersion lens for coupling nitrogen-vacancy (NV)-center photoluminescence to a collimated beam in air.  \textit{Inset}: Scanning electron microscope (SEM) image of fabricated metalens with closeup of etched diamond pillars. The scale bar corresponds to \SI{1}{\micro\meter}.  \textbf{b}, Bloch-mode effective index, $n_{\text{eff}}$, and corresponding optical pathlength difference, $\phi$, as a function of pillar diameter, $d$, at $\lambda = \SI{700}{\nano\meter}$. This map is used to create the lens pattern shown in \textbf{a}.}
\end{figure}
}

\def \MLDesign {
\begin{figure}[!t]
\includegraphics[width=\linewidth]{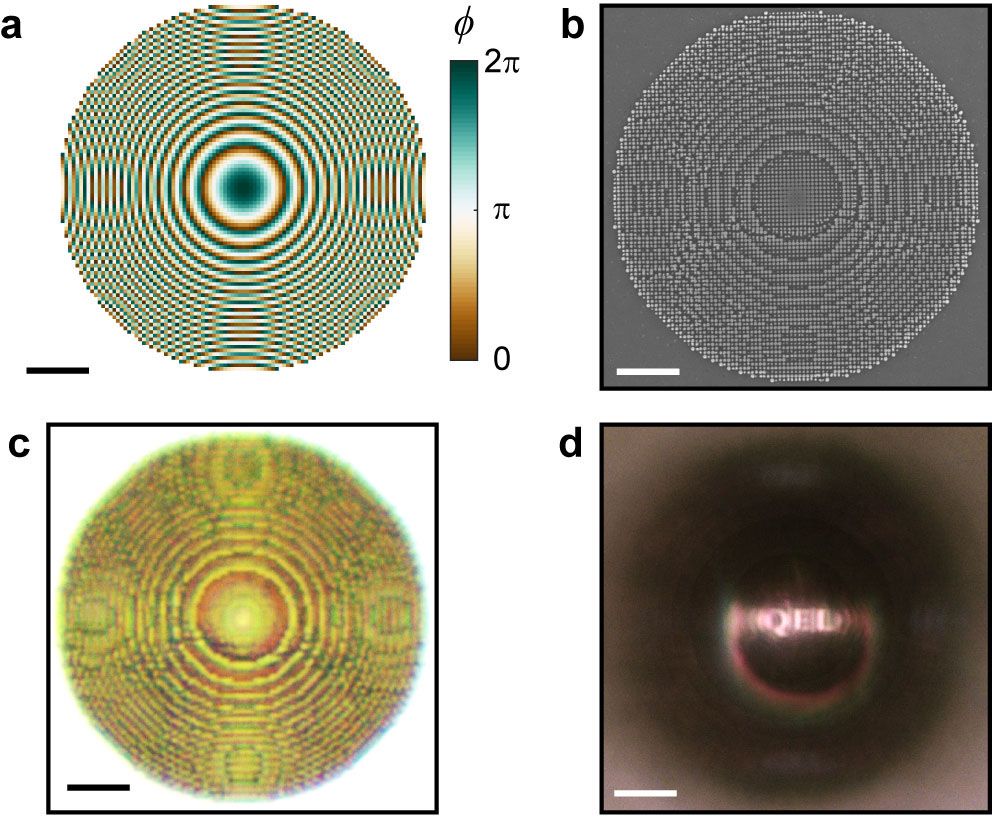}
\caption{\label{Fig:ML_design} \textbf{Metalens design and fabrication.} Top-down images of: \textbf{a}, the Fresnel phase profile used for the design; \textbf{b}, SEM image of pillar map, \textbf{c}, bright-field reflection optical micrograph of the metalens surface; and \textbf{d}, image of a macroscopic chromium shadow mask with the Quantum Engineering Lab logo, $\Braket{\text{Q}|\text{E}|\text{L}}$, formed through the metalens in a bright-field transmission microscope. All scale bars are \SI{5}{\micro\meter}.}
\end{figure}
}

\def \MLPerformance {
\begin{figure*}[!t]
\includegraphics[width=\textwidth]{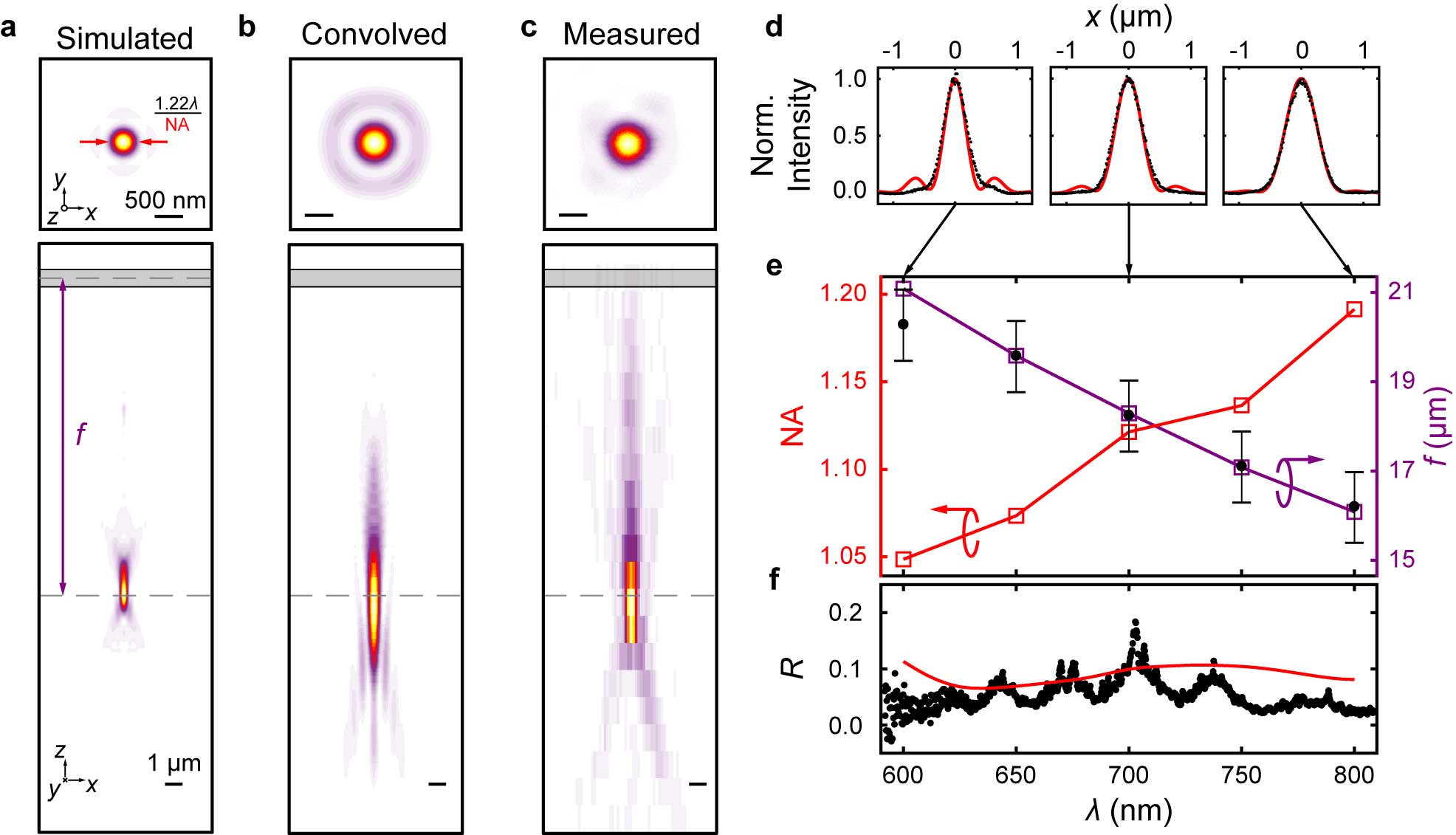}
\caption{\label{Fig:ML_performance}\textbf{Metalens performance.} \textbf{a-c} Transverse ($x-y$, top) and axial ($x-z$, bottom) cross-sections of the metalens focus at $\lambda = \SI{700}{\nano\meter}$ (\textbf{a}) simulated by a 3D finite-difference time-domain (FDTD) method, (\textbf{b}) calculated by coherently convolving the simulations in (\textbf{a}) with the microscope's point-spread function, and (\textbf{c}) measured using a confocal optical microscope. Grey boxes and dashed lines in axial cross-sections indicate the position of the metalens surface and focus, respectively. \textbf{d}, $x$-cross sections of the simulated metalens focus convolved with the microscope PSF (solid red curves) and measured metalens focus (points) at \SI{600}{\nano\meter}, \SI{700}{\nano\meter}, and \SI{800}{\nano\meter} wavelengths.  \textbf{e}, Metalens NA as a function of wavelength (open red squares), determined by fitting the simulated transverse focus cross-section with an Airy function, together with the measured (points with errorbars) and simulated (open purple squares) effective focal length. \textbf{f}, Simulated and measured metalens reflectivity.}
\end{figure*}
}

\def \MLwithNV {
\begin{figure*}[!t]
\includegraphics[width=\textwidth]{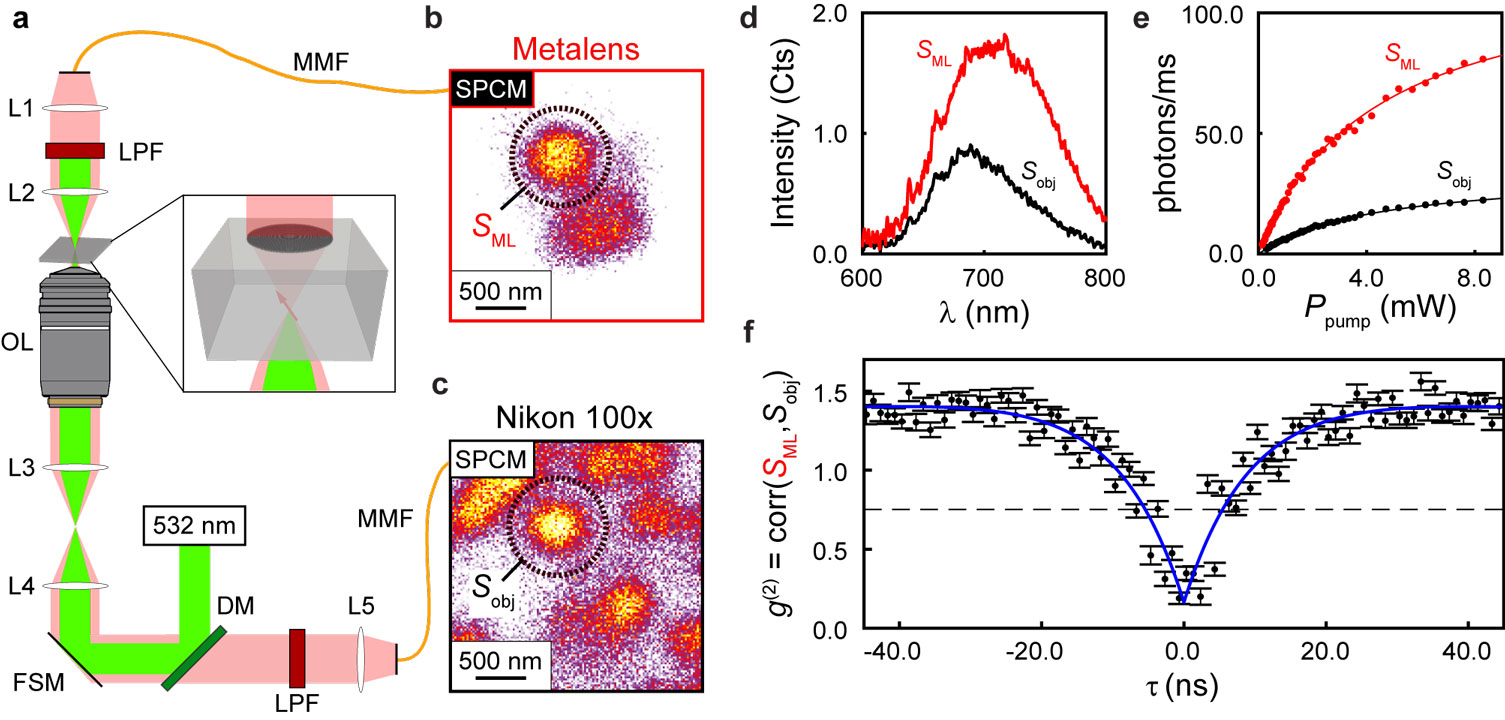}
\caption{\label{Fig:ML_with_NV} \textbf{Imaging an individual NV center |} \textbf{a}, Experimental setup. MMF = multimode fiber, LPF = longpass wavelength filter, OL = objective lens, FSM = fast steering mirror, DM = dichroic mirror, SPCM = single-photon counting module, L1-L5 are achromatic lenses. \textbf{b}, PL signal from the metalens when the \SI{532}{\nano\meter} pump beam is rastered.  \textbf{c}, confocal PL image from objective recorded simultaneously with (\textbf{b}).  \textbf{d}, PL spectra and \textbf{e}, saturation curves of the metalens and objective signals, $S_{\text{ML}}$ and $S_{\text{obj}}$, corresponding to the spot circled in (\textbf{b}) and (\textbf{c}), respectively. \textbf{f}, Intensity cross-correlation between $S_{\text{ML}}$ and $S_{\text{obj}}$, confirming that the spot measured in both images is an individual NV center. The dotted line represents the single-emitter threshold.  Measurements in (\textbf{d-f}) are background corrected.}
\end{figure*}
}

\title{Imaging a Nitrogen-Vacancy Center with a Diamond Immersion Metalens}

\author{Richard R. Grote}
\affiliation{Quantum Engineering Laboratory, Department of Electrical and Systems Engineering, University of Pennsylvania, 200 S. 33rd Street, Philadelphia, PA 19104, USA}

\author{Tzu-Yung Huang}
\affiliation{Quantum Engineering Laboratory, Department of Electrical and Systems Engineering, University of Pennsylvania, 200 S. 33rd Street, Philadelphia, PA 19104, USA}

\author{Sander A. Mann}
\altaffiliation[Current address: ]{Department of Electrical and Computer Engineering, University of Texas at Austin, Austin, TX 78701, USA}
\affiliation{Center for Nanophotonics, AMOLF, Science Park 104, 1098 XG Amsterdam, The Netherlands}

\author{David A. Hopper}
\affiliation{Quantum Engineering Laboratory, Department of Electrical and Systems Engineering, University of Pennsylvania, 200 S. 33rd Street, Philadelphia, PA 19104, USA}
\affiliation{Department of Physics and Astronomy, University of Pennsylvania, 209 S. 33rd Street, Philadelphia, PA 19104, USA}

\author{Annemarie L. Exarhos}
\altaffiliation[Current address: ]{Department of Physics, Lawrence University, Appleton, WI 54911, USA}
\affiliation{Quantum Engineering Laboratory, Department of Electrical and Systems Engineering, University of Pennsylvania, 200 S. 33rd Street, Philadelphia, PA 19104, USA}

\author{Gerald G. Lopez}
\affiliation{Singh Center for Nanotechnology, University of Pennsylvania, 3205 Walnut St., Philadelphia, PA 19104, USA}

\author{Erik C. Garnett}
\affiliation{Center for Nanophotonics, AMOLF, Science Park 104, 1098 XG Amsterdam, The Netherlands}

\author{Lee C. Bassett}
\email[Corresponding author: ]{lbassett@seas.upenn.edu}
\affiliation{Quantum Engineering Laboratory, Department of Electrical and Systems Engineering, University of Pennsylvania, 200 S. 33rd Street, Philadelphia, PA 19104, USA}

\date{\today}

\maketitle 
\Concept

\textbf{Solid-state quantum emitters have emerged as robust single-photon sources \cite{Aharonovich_NP_16} and addressable spins \cite{Gao_NP_15} | key components in rapidly developing quantum technologies for broadband magnetometry \cite{Rondin_RPP_14}, biological sensing \cite{Balasubramanian_COBC_14}, and quantum information science \cite{Awschalom_Science_13}. Performance in these applications, be it magnetometer sensitivity or quantum key generation rate, is limited by the number of photons detected. However, efficient collection of a quantum emitter's photoluminescence (PL) is challenging as its atomic scale necessitates diffraction-limited imaging with nanometer-precision alignment, oftentimes at cryogenic temperatures.  In this letter, we image an individual quantum emitter, an isolated nitrogen-vacancy (NV) center in diamond, using a dielectric metalens composed of subwavelength pillars etched into the diamond's surface (Fig.~\ref{Fig:Concept}\textbf{a}). The metalens eliminates the need for an objective by operating as a high-transmission-efficiency immersion lens with a numerical aperture (NA) greater than 1.0. This design provides a scalable approach for fiber coupling solid-state quantum emitters that will enable the development of deployable quantum devices.}

Beyond their atomic scale, the challenges associated with coupling to solid-state quantum emitters are exacerbated by the high refractive index of their host substrates. Diamond, for example, has a refractive index of $n_\mathrm{D}\sim 2.4$ at visible wavelengths, which traps photons emitted above $\theta_c\sim\SI{25}{\degree}$ of normal incidence at a planar air interface by total-internal reflection.  Furthermore, imaging through more than a few microns of diamond with a high-NA objective results in spherical aberrations that severely limit collection efficiency. While a number of nanophotonic structures have been investigated for increasing NV emission through Purcell enhancement\cite{Aharonovich_AOM_14,Loncar_MRS_13,Faraon_NJP_13,Schroder_JOSAB_16}, these devices require NVs positioned close to diamond surfaces, which degrades their spin\cite{Ofori_PRB_12} and optical properties\cite{Chu_NL_14}. For this reason, the typical approach to addressing single NVs in pristine bulk diamond is to mill or etch a hemispherical surface, known as a solid immersion lens (SIL), centered about an individual NV center. By achieving uniform optical path length and reflectance for rays emanating from the NV at all angles \cite{Castelletto_NJP_11}, SILs have removed the losses caused by total-internal-reflection and spherical aberration, enabling ground-breaking demonstrations in quantum optics such as the recent loop-hole-free violation of Bell's inequality\cite{Hensen_Nat_15}. However, a high-NA objective lens is still required to image a quantum emitter through a SIL. Thus, a cryostat that can accommodate a vacuum-compatible objective and associated optomechanics must be used, or the optical losses associated with imaging through a cryostat window must be accepted.  Neither option provides a clear route for packaging quantum emitters in a scalable fashion.

Since quantum emitters are point sources with relatively narrow emission spectra, the compound optical system of a microscope objective, which is designed for broadband imaging with a flat field-of-view, is not strictly necessary for efficient photon collection.  A more scalable approach would be to use flat optics, like the phase Fresnel lenses used to image trapped ions in ultra-high vacuum cryostats \cite{Jechow_OL_11}.  However, a flat optic on its own cannot compensate for the high-refractive index of a solid-state quantum emitter's host material. The ideal solution would be a flat optic fabricated at the air/diamond interface to form a planar immersion lens; such a design can be realized using the subwavelength elements of a \emph{metasurface}.

Metasurfaces have recently gained attention as they offer design flexibility for optical components with arbitrary phase responses \cite{Kildishev_Science_13,Yu_NM_14}. In particular, dielectric metalenses \cite{Lalanne_LPR_17,Genevet_Optica_17,Khorasaninejad_Science_16}, diffractive optics \cite{Lalanne_LPR_17,Lee_JOA_02}, and high-contrast gratings \cite{Chang_AOP_12,Vo_PTL_14} comprised of high-refractive-index dielectric elements such as TiO$_2$ and amorphous silicon have been demonstrated with high transmission efficiency and diffraction-limited focusing. While spherical and chromatic aberrations limit the field-of-view of single-element dielectric metalenses as compared to aberration-corrected multi-lens objectives \cite{Lalanne_LPR_17}, they are ideally suited for collimating emission from a point source over a narrow wavelength range.

Building on these advances, we leverage diamond's high refractive index to image an individual NV center located $\sim$\SI{20}{\micro\meter} below a $\sim$\SI{28}{\micro\meter}-diameter metalens fabricated on the surface of a single-crystal substrate. We demonstrate a transmission efficiency $>$88\% and $\mathrm{NA}>1.0$, and use the metalens to couple NV PL into a fiber with a background-subtracted saturation count rate of $\sim$122~photons/ms.  This marks the first step in designing and fabricating arbitrary metasurfaces for controlling emission from quantum emitters using only top-down fabrication techniques and provides a clear pathway to packaging quantum devices by eliminating the need for an objective.

\MLDesign
The metalens is fabricated using electron beam lithography and O$_2$-based dry etching to produce the subwavelength pillars seen in the inset of Fig.~\ref{Fig:Concept}\textbf{a}. These pillars approximate a desired continuous phase profile, $\phi(x,y)$, on a square grid, by mapping the pillar diameter, $d$, to the effective refractive index, $n_{\text{eff}}$, of the lowest-order Bloch-mode supported by the pillar (Fig.~\ref{Fig:Concept}\textbf{b}). We use a Fresnel lens phase profile in conjunction with Fig.~\ref{Fig:Concept}\textbf{b} to assign a pillar diameter to each grid point.  The discretized phase profile for a focal length $f = \SI{20}{\micro\meter}$ at $\lambda = \SI{700}{\nano\meter}$ is shown in Fig.~\ref{Fig:ML_design}\textbf{a}, with a corresponding SEM image of the fabricated structure shown in Fig.~\ref{Fig:ML_design}\textbf{b}. The pillars are inherently anti-reflective (see supporting information), which is evidenced by the bright-field reflection microscope image of the metalens surface shown in Fig.~\ref{Fig:ML_design}\textbf{c}. To demonstrate that the structure operates as a lens, in Fig.~\ref{Fig:ML_design}\textbf{d} we use a transmission microscope to form an image through the metalens of a chromium shadow mask below the diamond (see supporting information).

\MLPerformance
We characterize the metalens using a combination of three-dimensional full-field electromagnetic simulations and confocal microscopy. When illuminated by a plane wave in air, the metalens forms a focused spot in the diamond as shown by the simulations in Fig.~\ref{Fig:ML_performance}\textbf{a}.  We measure the focused field distribution by illuminating the metalens from above with a collimated laser beam, while imaging the transmitted field using a scanning confocal microscope with an oil immersion objective situated below the diamond. Spherical aberrations caused by imaging through the $\sim\SI{150}{\micro\meter}$-thick diamond plate limit the resolution of these measurements, resulting in a focus spot that appears larger than the physical field profile inside of the diamond. To accurately compare simulations and measurements, we numerically model the microscope's point-spread function (see Methods) and coherently convolve it with the simulated focus spot to predict the measured transverse and axial field profiles (Fig~\ref{Fig:ML_performance}\textbf{b}). These predicted field profiles show excellent agreement with the measurements in Fig.~\ref{Fig:ML_performance}\textbf{c}, as evidenced by the cross-sections shown in Fig.~\ref{Fig:ML_performance}\textbf{d} for multiple wavelengths. Similar agreement is observed between simulations and measurements of the metalens focal length, $f$ (Fig.~\ref{Fig:ML_performance}\textbf{e}).

The widths of the simulated field profiles are used to determine the NA of the metalens as a function of wavelength (Fig.~\ref{Fig:ML_performance}\textbf{e}), showing $\text{NA}>1.0$ across all wavelengths of the NV's full emission spectrum.  It is worth noting that the high NA of our metalens is achieved by using diamond as an immersion medium, whereas previous high-NA metalenses have relied on diffraction far from the the optical axis to focus wide angles \cite{Lee_JOA_02,Khorasaninejad_Science_16}. This implies that the NA of our diamond design metalens could be substantially increased to values approaching the maximum $\text{NA}=n_\mathrm{D}=2.4$ by using higher-order diffraction to focus larger angles. In addition to exhibiting a high NA, the low reflectivity seen in Fig.~\ref{Fig:ML_design}\textbf{c} is quantified by the simulation and measurement to be below 11.5\% (Fig.~\ref{Fig:ML_performance}\textbf{f}).


\MLwithNV
To image an NV center with the metalens, we focus a \SI{532}{\nano\meter} pump beam through the backside of the substrate using an oil immersion objective (Fig.~\ref{Fig:ML_with_NV}\textbf{a}). The confocal collection/excitation volume of the objective is axially positioned in the plane of the metalens focus, and is rastered using a fast steering mirror (FSM).  NV PL at each scan position is simultaneously measured by two fiber-coupled single-photon counting modules (SPCMs): one aligned to the metalens, and the other aligned to the confocal path through the objective.  The counts collected by the SPCMs at each point of the FSM raster scan form the images shown in Fig.~\ref{Fig:ML_with_NV}\textbf{b,c}. The lenses in the metalens path (L1,L2 in Fig.~\ref{Fig:ML_with_NV}\textbf{a}) re-collimate the diverging metalens output beam so that a \SI{568}{\nano\meter} long-pass filter (LPF) can be inserted to block the pump beam.  Alternatively, the metalens output can be coupled directly into a fiber, if the pump beam is removed using a different excitation geometry or a commercially-available multilayer-coated fiber tip (Omega Optical, Inc., for example).

Figures~\ref{Fig:ML_with_NV}\textbf{b} and~\ref{Fig:ML_with_NV}\textbf{c} both exhibit a bright spot at the same lateral position, denoted by the black dashed circles. We fix the FSM position at the center of this spot and measure the PL signals ($S_{\text{ML,obj}}$) through the metalens and objective paths, respectively. Background signals are separately recorded from a position off the spot but within the metalens field of view (see supporting information). The background-subtracted spectra of both paths (Fig.~\ref{Fig:ML_with_NV}\textbf{d}) clearly exhibit the NV center's zero-phonon line at \SI{637}{\nano\meter} and characteristic phonon side band.  Background-subtracted PL saturation curves (Fig.~\ref{Fig:ML_with_NV}\textbf{e}) display saturation count rates of $121.7\pm2.2$~photons/ms and $33.5\pm0.6$~photons/ms when measured through the metalens and objective, respectively. The ratio of saturation count rates provides an estimate for the metalens collection efficiency, further indicating that the metalens has NA~$>1.0$ (see Methods).  Finally, we measure the second-order cross-correlation function, $g^{(2)}(\tau)$, between both paths. The background-corrected $g^{(2)}$ measurements (Fig.~\ref{Fig:ML_with_NV}\textbf{f}) exhibit the characteristic antibunching dip and short-delay bunching of a single NV center, clearly demonstrating that the spots in Fig.~\ref{Fig:ML_with_NV}\textbf{b,c} are indeed the same single-photon emitter.

The diamond immersion metalens lays the foundation for packaging quantum emitters in high-refractive-index substrates, as it has the potential to significantly improve emitter collection efficiency and simplify experiments by replacing the objective/SIL combination typically used for imaging quantum emitters in a cryostat. This approach can be directly applied to other quantum-emitter-host materials including silicon carbide, III-V semiconductors, and oxides. Leveraging the structure's high transmission efficiency and scalable top-down fabrication, other metasurface phase profiles can be explored to further increase the metalens NA by designing for large-angle diffraction \cite{Lee_JOA_02}, co-focusing pump and PL wavelengths \cite{Wang_NL_16,Ribot_AOM_13}, shaping emission from quantum emitter ensembles \cite{Lawrence_JAP_12}, and as a means for compensating mismatches between emitters and surface orientations \cite{Backlund_NP_16}. In addition, this type of metasurface could be incorporated with nanophotonic structures for Purcell enhancement, for example to collimate the output of a chirped surface grating structure through the backside of the diamond\cite{Zheng_arXiv_17}, or to extend the cavity length of a fiber-based resonator cavity \cite{Bogdanovic_APL_17}. Dielectric metasurface design will lead to compact, fiber-coupled single-photon sources and quantum memories, with other potential applications to diffractive optics for space \cite{Nikolajeff_OED_13} and Raman lasers \cite{Mildren_OED_Ch8_13}.

\section*{Acknowledgements}
This work was supported by an NSF CAREER grant (ECCS-1553511), the University Research Foundation, and the Singh Center for Nanotechnology at the University of Pennsylvania, a member of the National Nanotechnology Coordinated Infrastructure (NNCI), which is supported by the National Science Foundation (Grant ECCS-1542153). S.A.M. and E.C.G. were supported by the Netherlands Organisation for Scientific Research (NWO) and the European Research Council under the European Union’s Seventh Framework Programme ((FP/2007-2013)/ERC grant agreement no. 337328, “Nano-EnabledPV”).

\section*{Author contributions}

R. R. G. and T.-Y. H. contributed equally to this work. R. R. G. and L. C. B. conceived of the project. R. R. G., S. A. M., and E. C. G. performed the design and simulations; R. R. G. and G. G. L. fabricated the metalens; R. R. G., T.-Y. H., D. A. H., A. L. E., and L. C. B. performed the measurements and analysis.  All authors contributed to writing the manuscript.

\section*{Methods}

\noindent \textbf{Design.} The metalens was designed using the procedure devised by \citeauthor{Lalanne_JOSAA_99} for TiO$_2$ deposited on glass \cite{Lalanne_JOSAA_99}. The procedure was carried out as follows: First, the Bloch-mode effective index, $n_{\text{eff}}$, was calculated as a function of pillar diameter (Fig.~\ref{Fig:Concept}\textbf{b}) on a subwavelength grid.  The grid-pitch, $\Lambda$, was chosen to be just below the onset of first order diffraction, $\Lambda \leq \frac{\lambda}{n_\text{D}} = \SI{291}{\nano\meter}$ at $\lambda = \SI{700}{\nano\meter}$, which was rounded up to $\Lambda = \SI{300}{\nano\meter}$. The pillar height was chosen to be $h = \SI{1.0}{\micro\meter}$ and the minimum pillar diameter was set to $d_{\text{min}}=\SI{100}{\nano\meter}$ to ensure compatibility with our fabrication process.  The maximum pillar diameter, $d_{\text{max}}$, was then found by determining the $n_{\text{eff}}$ required to achieve an optical pathlength increase of $2\pi$ relative to the minimum pillar diameter:

\begin{equation}
n_{\text{eff}}\left(d_{\text{max}}\right) = \frac{\lambda}{h} + n_{\text{eff}}\left(d_{\text{min}}\right).
\label{eqn:PhiMax}
\end{equation}

\noindent The corresponding $d_{\text{max}}$ is found from the dispersion curve in Fig.~\ref{Fig:Concept}\textbf{b}. The minimum and maximum pillar diameters are indicated in Fig.~\ref{Fig:Concept}\textbf{b} (black dashed lines) along with the their relative optical pathlengths (red dashed lines).

The Fresnel phase profile in Fig.~\ref{Fig:ML_design}\textbf{a} was calculated by $\phi = n_
\mathrm{D}k_0 \left(f-\sqrt{f^2 + x^2 + y^2}\right)$, with 93 grid points for a diameter of \SI{27.9}{\micro\meter} measured by the grid edges at the maximum widths along the Cartesian design dimensions.  The symmetry of this structure ensures polarization independent focusing, which has been shown for similar designs using TiO$_2$ deposited on glass \cite{Khorasaninejad_NL_16}.

\vspace{.5cm}
\noindent \textbf{Fabrication.} The metalens was fabricated on $\SI{3.0}{\milli\meter}\times\SI{3.0}{\milli\meter}\times\SI{0.15}{\milli\meter}$ double-side polished high-pressure/high-temperature (HPHT)-grown single-crystal diamond (Applied Diamond, Inc.).  The diamond surface was cleaned in \SI{90}{\celsius} Nano-Strip (a stabilized mixture of sulfuric acid and hydrogen peroxide, Cynaktec KMB 210034) for \SI{30}{\minute}, followed by a \SI{10}{\minute} plasma clean in a barrel asher with 40~sccm O$_2$ and \SI{300}{\watt} RF power. The metalens pattern was proximity effect corrected (see supporting information) and written in hydrogen silsesquioxane (HSQ, Dow Corning, Fox-16) using a 50 keV electron beam lithography tool (Elionix, ELS-7500EX).  Prior to spin-coating HSQ, a \SI{7}{\nano\meter} adhesion layer of SiO$_2$ was deposited on the diamond surface by electron beam evaporation to promote adhesion.  After exposure, the pattern was developed in a mixture of \SI{200}{\milli\liter} deionized water with \SI{8}{\gram} of sodium chloride and \SI{2}{\gram} of sodium hydroxide \cite{Yang_JVSTB_07}.  Our e-beam lithography process for HSQ on diamond can be found in ref.~\onlinecite{Grote_SC_16}.  A reactive ion etch (RIE, Oxford Instruments, Plasma lab 80) was used to remove the SiO$_2$ adhesion layer and to transfer the HSQ pattern into the diamond surface. The SiO$_2$ adhesion layer was removed by a \SI{1}{\minute} CF$_4$ reactive ion etch \cite{Metzler_SC_16}, followed by a \SI{23}{\minute} O$_2$ RIE etch with a flow rate of 40~sccm, a chamber pressure of \SI{75}{\milli\torr}, and an RF power of \SI{200}{\watt} to form the diamond pillars.  Finally, the HSQ hardmask was removed using buffered-oxide etch.

\vspace{.5cm}
\noindent \textbf{Simulations.}
Calculations of $n_{\text{eff}}$, $\phi$ (Fig.~\ref{Fig:Concept}\textbf{b}, left and right axes, respectively), and pillar transmission efficiency (supporting information) were performed using 3D rigorous coupled-wave analysis (RCWA) based on the method developed by \citeauthor{Rumpf_PIERS_11}\cite{Rumpf_PIERS_11}. The effective index of the pillars was calculated by solving for the eigenvalues of Maxwell's equations with the $z$-invariant refractive index profile of the pillar cross-section in a $\SI{300}{\nano\meter}\times\SI{300}{\nano\meter}$ square unit cell at $\lambda = \SI{700}{\nano\meter}$.  The eigenproblem was defined in a truncated planewave basis using $25\times25$ planewaves, with implicit periodic boundary conditions. Following these calculations, the pillar height was set to \SI{1.0}{\micro\meter} with air above and homogeneous diamond below, and the complex amplitude transmission coefficient, $t$, of a normal incidence planewave from air is calculated as a function of pillar diameter.  The right axis of Fig.~\ref{Fig:Concept}\textbf{b} was found by $\phi(d) = \angle t(d)$.

The focused spot in Fig.~\ref{Fig:ML_performance}\textbf{a} was calculated using 3D finite-difference time-domain simulations (FDTD, Lumerical Solutions, Inc.). The \SI{27.9}{\micro\meter}-diameter metalens is contained in a $\SI{28.1}{\micro\meter}\times\SI{28.1}{\micro\meter}\times\SI{22.25}{\micro\meter}$ total-field/scattered-field (TFSF) excitation source to reduce artifacts caused by launching a planewave into a finite structure. Perfectly matched layers (PMLs) were used as boundary conditions \SI{0.5}{\micro\meter} away from the TFSF source. The simulation mesh in the pillars was set to $\SI{10}{\nano\meter}\times\SI{10}{\nano\meter}\times \SI{10}{\nano\meter}$, increasing gradually to \SI{50}{\nano\meter} along the propagation ($\hat{z}$)-direction into the diamond. Diamond is modeled with a non-dispersive refractive index, $n_D=2.4$. An $x$-polarized planewave pulse ($\omega_0 \approx 2\pi\times\SI{440}{\tera\hertz},\Delta \omega \approx 2\pi\times\SI{125}{\tera\hertz}$) is launched from air toward the metalens surface. Steady-state spatial electric field distributions, $\vec{E}(\vec{r})$, at five wavelengths ranging from \SIrange{600}{800}{\nano\meter} were stored, and the spatial fields at $\lambda = \SI{700}{\nano\meter}$ are plotted as transverse ($|\vec{E}(z=f)|^2$) and axial ($|\vec{E}(y=0)|^2$) intensity distributions in Fig.~\ref{Fig:ML_performance}\textbf{a}. The focal length, $f_{\text{ML}}$, at each wavelength (Fig.~\ref{Fig:ML_performance}\textbf{e}) was determined by finding the grid point in the simulation cell where $|\vec{E}|^2$ is maximum. The spatial distribution of the steady-state field amplitude, $E_x(\vec{r})$, in Fig.~\ref{Fig:Concept}\textbf{a} was simulated by removing the TFSF source and placing an $\hat{x}$-oriented dipole current source at the metalens focus position with a wavelength of \SI{700}{\nano\meter}.  The reflection spectrum (Fig.~\ref{Fig:ML_performance}\textbf{f}) was calculated by integrating the time-averaged Poynting vector, $S_z = -\frac{1}{2}\text{Re}\left\{\vec{E}\times\vec{H}^*\right\}\cdot \hat{z}$, over a $\SI{30}{\micro\meter}\times\SI{30}{\micro\meter}$ surface, \SI{0.1}{\micro\meter} above and \SI{0.4}{\micro\meter} below the metalens within the TFSF source volume. The simulation volume was reduced to $\SI{31}{\micro\meter}\times\SI{31}{\micro\meter}\times\SI{2}{\micro\meter}$ and the number of wavelength points was increased to 41 for these simulations.

The images in Fig.~\ref{Fig:ML_performance}\textbf{b} represent the optical intensity, $I$, collected by a detector at a  focus position in the sample, $\vec{r}_{\text{image}}$, defined by the FSM in the transverse directions and by the sample stage in the axial direction: $\vec{r}_{\text{image}} = x_{\text{FSM}}\cdot\hat{x} + y_{\text{FSM}}\cdot\hat{y} + z_{\text{piezo}}\cdot\hat{z}$.  These images are produced by coherently convolving the FDTD-calculated steady-state fields, $\vec{E}(\vec{r})$, with the point-spread function (PSF) of the microscope, which is modeled by numerically evaluating the diffraction integrals, $I_0,I_1,I_2$, that define the dyadic Green's function of a high-NA optical system \cite{Novotny_12}:

\begin{align}
\mathbf{G}&(\vec{r}_{\text{image}},\vec{r},\lambda) = \left[\begin{array}{ccc}
		G_{xx} & G_{xy} & G_{xz} \\
		G_{yx} & G_{yy} & G_{yz} \\
		0 & 0 & 0
\end{array}\right] \nonumber\\
=& \left[\begin{array}{ccc}
		I_0 + I_2\cos2\phi & I_2 \sin2\phi & -2jI_1\cos\phi \\
		I_2\sin2\phi & I_0 - I_2\cos2\phi & -2jI_1\sin\phi \\
		0 & 0 & 0
\end{array}\right],
\label{Eqn:GreensFunc}
\end{align}

\noindent with the inclusion of an aberration function that accounts for the optical pathlength difference introduced by imaging through a media with mismatched refractive indices \cite{Sheppard_JM_97} ($n_{\text{oil}} = 1.518$ and $n_{\text{D}} = 2.4$ for our measurement setup).  We assume an infinitesimal pinhole, which is consistent with our imaging system being below the confocal condition (see supporting information). Using Eqn.~(\ref{Eqn:GreensFunc}), the image formed by our microscope is modeled in the following manner (see supporting information):

\begin{align}
I(\vec{r}_{\text{image}}) &= |G_{xx}*E_x + G_{xy}*E_y + G_{xz}*E_z|^2 \nonumber \\
 &+ |G_{yx}*E_x + G_{yy}*E_y + G_{yz}*E_z|^2
\label{Eqn:coherentConvlution}
\end{align}

\noindent where $*$ denotes a three-dimensional spatial convolution.  The transverse, $I(z_{\text{piezo}}=f)$, and axial, $I(y_{\text{FSM}}=0)$, image intensity distributions at $\lambda = \SI{700}{\nano\meter}$ are shown in Fig.~3\textbf{b}, and cross-sections, $I(y_{\text{FSM}}=0,z_{\text{piezo}}=f)$, at $\lambda = \SI{600}{\nano\meter},\SI{700}{\nano\meter},\SI{800}{\nano\meter}$ are plotted in Fig.~\ref{Fig:ML_performance}\textbf{d} (red curves).  Transverse profiles at five wavelengths ranging from $\lambda = \SIrange{600}{800}{\nano\meter}$ are plotted in the supporting information.

\vspace{.5cm}
\noindent \textbf{Experimental.}
Measurements of the metalens were carried out with a custom-built confocal microscope, comprised of an oil immersion objective with adjustable iris (Nikon Plan Fluor x100/0.5-1.30) and an inverted optical microscope (Nikon Eclipse TE200) with a $\hat{z}$-axis piezo stage (Thorlabs MZS500-E) as well as a scanning stage for the $\hat{x}$- and $\hat{y}$-axis (Thorlabs MLS203-1).  The diamond host substrate was fixed to a microscope coverslip (Fisher Scientific 12-548-C) using immersion oil (Nikon type N) with the patterned surface facing upwards. A combination of $\SI{30}{\milli\meter}$ cage system and SM1-thread components (Thorlabs) were used to create a fiber-coupled optical path above the stage of the inverted microscope.  This configuration allowed for simultaneous excitation and measurement of the metalens from air (fiber-coupled path) or through diamond (objective path).  The objective path was routed outside of the microscope body so that laser-scanning confocal excitation and collection optics could be added.  A $4f$ relay-lens-system consisting of two achromatic doublet lenses (Newport, $\SI{25.4}{\milli\meter}\times\SI{150}{\milli\meter}$ focal length, PAC058AR.14) is used to align the back aperture of the objective to a fast-steering mirror (FSM, Optics in motion, OIM101), which is used to raster the diffraction-limited confocal volume in the transverse $x-y$ plane of the objective space. A \SI{560}{\nano\meter} long-pass dichroic mirror (Semrock, BrightLine FF560-FDi01) placed after the FSM was used to couple a \SI{532}{\nano\meter} excitation laser (Coherent, Compass 315M-150) into the objective, while wavelengths above \SI{560}{\nano\meter} pass through the dichroic mirror and are focused into a \SI{25}{\micro\meter}-core, 0.1~NA, multimode fiber (Thorlabs M67L01) that can be connected to a single-photon counting module (Excelitas, SPCM-AQRH-14-FC) or a spectrometer (Princeton Instruments IsoPlane-160, \SI{750}{\nano\meter} blaze wavelength with 1200 G/mm) with a thermoelectrically-cooled CCD (Princeton Instruments PIXIS 100BX). Computer control of the FSM and counting the electrical output of the SPCM are achieved using a data acquisition card (DAQ, National Instruments PCIe-6323).

For the characterization measurements presented in Fig.~3, a broadband supercontinuum source (Fianium WhiteLase SC400) was coupled into a single-mode fiber (Thorlabs P1-630AR-2), which was used to illuminate the metalens from the fiber-coupled path of our microscope. A $f = \SI{2.0}{\milli\meter}$ collimating lens (Thorlabs CFC-2X-A) was used to create a \SI{380}{\micro\meter} diameter Gaussian beam that emulates the planewave source used in our FDTD simulations. The excitation wavelength is set by passing the supercontinuum beam through a set of linear variable short-pass (Delta Optical Thin Film, LF102474) and long-pass filters (Delta Optical Thin Film LF102475) prior to fiber-coupling, which can be adjusted to filter out a single wavelength with $<\SI{8}{\nano\meter}$ bandwidth or be removed completely for broadband excitation.  The transverse profile and cross-sections in Fig.~\ref{Fig:ML_performance}\textbf{c,d} were measured by filtering the supercontinuum source to a single wavelength and rastering the FSM while collecting counts in the SPCM connected to the confocal path at each scan position.  This process is repeated for a series of $z$-stage positions to measure the axial profile, which is shown in Fig.~\ref{Fig:ML_performance}\textbf{c} at $\lambda = \SI{700}{\nano\meter}$ and was used to find the metalens focal length as a function of wavelength in Fig.~\ref{Fig:ML_performance}\textbf{e}.  For reflection measurements (Fig.~\ref{Fig:ML_performance}\textbf{f}) a $f = \SI{15}{\milli\meter}$ achromatic lens (Thorlabs AC064-015-B) is used to focus the collimated excitation beam to a $\sim\SI{30}{\micro\meter}$-diameter spot at the top surface of the diamond.  A beamspliter cube (Thorlabs BS014) was added between the collimating and focusing lenses so that reflected light could be focused into a \SI{200}{\micro\meter}-core MMF (Thorlabs, M25L01) that is coupled to a spectrometer (Thorlabs CCS100) using a $f = \SI{100}{\milli\meter}$ achromatic doublet lens (Newport, PAC052AR.14).


In Fig.~\ref{Fig:ML_with_NV}, the fiber-coupled path was used to image a single NV center through the metalens, as shown in Fig.~\ref{Fig:ML_with_NV}\textbf{a}.
This was achieved with two achromatic doublet lenses (L1 \& L2) with focal lengths of $f = \SI{13}{\milli\meter}$ and $f = \SI{15}{\milli\meter}$ (Thorlabs AC064-013/015-B), respectively, aligned to a \SI{25}{\micro\meter}-core, 0.1~NA, multimode fiber (Thorlabs M67L01).
The multimode fiber was then connected to a second SPCM (Excelitas, SPCM-AQRH-14-FC), allowing for simultaneous PL collection from both the fiber-coupled and objective paths while scanning the excitation source.
The long-pass filters (LPF) in both collection lines consisted of a \SI{532}{\nano\meter} and a \SI{568}{\nano\meter} long-pass filters (Semrock, EdgeBasic BLP01-532R, EdgeBasic BLP01-568R) for spectra measurements, with an additional $\SI{650}{\nano\meter}$ long-pass filter (Thorlabs, FEL0650) in both paths to improve the signal-to-background for PL, saturation, and cross-correlation measurements.
The outputs of both SPCMs were connected to a time-correlated single-photon counting card (TCSPC, PicoQuant, PicoHarp 300) to collect photon arrival-time data that was used to calculate cross-correlation functions (Fig.~\ref{Fig:ML_with_NV}\textbf{f}). Background spectra and saturation curves were measured at a transverse scan position away from the NV, but still within the field-of-view of the metalens, and were subtracted from measurements taken on the NV.  This process was also used to determine the background for correcting cross-correlation data by interleaving 40 measurements off the NV with 40 measurements taken on the NV, each with a \SI{5}{\minute} acquisition time. Further details on background-subtraction of the measurements in Fig.~\ref{Fig:ML_with_NV} are given in the supporting information.

\vspace{.5cm}
\noindent{\textbf{Analysis}.} The NA of the metalens, NA$_{\text{ML}}$, plotted in Fig.~\ref{Fig:ML_performance}\textbf{e} is calculated by fitting the simulated transverse focus spot at each wavelength to the paraxial point-spread function of an ideal lens, an Airy disk \cite{Novotny_12},

\begin{equation}
I = \left|\frac{2J_1\left(\text{NA}_{\text{ML}}k_0r\right)}{\text{NA}_{\text{ML}}k_0r}\right|^2,
\label{Eqn:Airy}
\end{equation}

\noindent where $k_0 = 2\pi/\lambda$ is the free space wavenumber and $r = \sqrt{x^2 + y^2}$ is the radial coordinate in the focal plane.  Fits are performed using non-linear least squares curve fitting (MATLAB function \textbf{lsqcurvefit}). The entrance pupil, $D$, of the metalens can be calculated by geometry using NA$_{\text{ML}}$ and $f_{\text{ML}}$:

\begin{equation}
D = 2f_{\text{ML}}(\lambda)\tan\left[\sin^{-1}\left(\frac{\text{NA}_{\text{ML}}}{n_{\text{D}}}\right)\right].
\label{Eqn:EntrancePupil}
\end{equation}

\noindent Using Fig.~\ref{Fig:ML_performance}\textbf{e} along with eqn.~(\ref{Eqn:EntrancePupil}), we find that $D = \SI{19.3}{\micro\meter}$, which is smaller than the physical \SI{27.9}{\micro\meter} diameter of the metalens.  This indicates a maximum collection angle inside the diamond of $\theta_{\text{max}} = \sin^{-1}\left(\frac{\text{NA}_{\text{ML}}}{n_{\text{D}}}\right) = \SI{27.8}{\degree}$.  Despite this limited collection angle, Fig.~\ref{Fig:ML_performance}\textbf{e} clearly illustrates NA$_{\text{ML}} > 1.0$, which can be increased by using diffractive designs for larger angles.

The focal length of the metalens in Fig.~\ref{Fig:ML_performance}\textbf{e} was determined by measuring the distance between the metalens surface and the focused spot formed below the metalens using the piezo stage of the microscope.  The distance traversed by the piezo stage is then scaled by a factor of $\approx \frac{n_{\text{D}}}{n_{\text{oil}}}$ to compensate for distortions caused by imaging through diamond \cite{Visser_Scanning_94}.  Further details are given in the supporting information.

The reflectance spectrum in Fig.~\ref{Fig:ML_performance}\textbf{f} was normalized using measurements of the reflected optical power measured with the fiber-coupled path aligned to the metalens, $P_{\text{ML}}(\lambda)$, and off the metalens on a planar region of the diamond surface, $P_{\text{surface}}(\lambda)$, using the following expression:

\begin{equation}
R_{\text{ML}}(\lambda) = \frac{P_{\text{ML}}(\lambda)}{P_{\text{surface}}(\lambda)}R_{\text{surface}},
\label{Eqn:RefNorm}
\end{equation}

\noindent where $R_{\text{surface}} = \frac{P_{\text{surface}}(\lambda)}{P_{\text{in}}(\lambda)}$ is the reflectance of an air/diamond interface and is calculated using Fresnel coefficients to be $17\%$ at normal incidence.  The ripples in Fig.~\ref{Fig:ML_performance}\textbf{f} are due to ghosting from the beam splitter cube used to collect the reflected signal (see supporting information).  The measured reflectance spectrum is slightly lower than the simulated spectrum (both plotted in Fig.~\ref{Fig:ML_performance}\textbf{f}).  The source of the discrepancy is believed to be due to the NA of our top collection optics.  The simulations represent the reflected light over all angles (specular and scattered), while our collection optics only cover a limited range of angles.

The saturation curves in Fig.~\ref{Fig:ML_with_NV}\textbf{e} were fit with the following equation:
\begin{equation}
C = \frac{C_{\text{sat}}}{1 + \frac{P_{\text{sat}}}{P_{\text{pump}}}},
\label{Eqn:Saturation}
\end{equation}
\noindent using non-linear least squares curve fitting (MATLAB function \textbf{lsqcurvefit}), resulting in saturation count rates of $C^{\text{ML}}_{\text{sat}} = 121.7\pm2.2$~photons/ms and $C^{\text{obj}}_{\text{sat}} = 33.5\pm0.6$~photons/ms for the metalens signal, $S_{\text{ML}}$, and confocal signal, $S_{\text{obj}}$, respectively. The saturation power was $P_{\text{sat}} = 4.3 \pm 0.1~\si{\milli\watt}$ in both paths, since they are both pumped by the same excitation beam.

The collection efficiency as a function of numerical aperture can be estimated as \cite{Castelletto_NJP_11}:
\begin{align}
\eta =& \frac{1}{32}\Bigg[ 15 \left(1-\sqrt{1-\left(\frac{\text{NA}}{n_{\text{D}}}\right)^2}\right) \nonumber \\
&+  \left(1 - \cos\left[3 \sin^{-1}\left(\frac{\text{NA}}{n_{\text{D}}}\right) \right]\right)\Bigg].
\label{Eqn:CollectionEfficiency}
\end{align}
Assuming that the excitation and collection paths have similar transmission efficiencies, the ratio of collection efficiencies from both paths is equal to the ratio of saturation count rates, $\frac{\eta_{\text{ML}}}{\eta_{\text{obj}}} = \frac{C^{\text{ML}}_{\text{sat}}}{C^{\text{obj}}_{\text{sat}}}$. Using a numerical aperture of NA$_{\text{obj}} = 0.75$ for the confocal collection path, the metalens is estimated to have NA$_{\text{ML}} \approx 1.4$. If instead we assume that the ratio of the collection efficiencies is proportional to the ratio of the integrated spectra in Fig.~\ref{Fig:ML_with_NV}\textbf{d}, we find that NA$_{\text{ML}} = 1.16$.  Discrepancies in these values arise from differences in the collection efficiency of both paths caused by the confocal aperture and optical components in the path.  However, this rough calculation provides strong evidence that NA$_{\text{ML}}>1.0$.

Background-correction of the cross-correlation data in Fig.~\ref{Fig:ML_with_NV}\textbf{f} was performed using the following relationship\citep{Brouri_OL_00}:

\begin{equation}
g^{(2)}_{\text{bc}}(\tau)
= \dfrac{g^{(2)}(\tau) - (1 - \rho^2)}{\rho^2}
\label{Equ:gCorrected}
\end{equation}

\noindent where $g^{(2)}(\tau)$ is the measured second-order correlation function and $\rho = 0.26\pm0.01$ is the total signal-to-background ratio determined by 40 repeated measurements. After background correction, $g^{(2)}_{\text{bc}}(\tau)$ is fit with the following expression:

\begin{align}
g^{(2)}_{\text{bc}}(\tau) &= 1 - Ae^{-\tfrac{\abs{t-t_0}}{\tau_1}} + Ce^{-\tfrac{\abs{t-t_0}}{\tau_2}},
\label{Equ:corrFunc}
\end{align}

\noindent which corresponds to the the approximation of the NV center as a 3-level structure\cite{Kitson_PRA_98}. The fit coefficients are as follows: $A = 1.31\pm0.05, C = 0.48\pm0.02, \tau_1 = 8.82\pm\SI{0.05}{\nano\second}, \tau_2 = 220.89\pm\SI{9.28}{\nano\second}$.  Further details are given in the supporting information.

\section*{References}
\bibliography{Metalens_NP_17}

\end{document}


\title{Supplementary Information: Imaging a Nitrogen-Vacancy Center with a Diamond Immersion Metalens}

\author{Richard R. Grote}
\affiliation{Quantum Engineering Laboratory, Department of Electrical and Systems Engineering, University of Pennsylvania, 200 S. 33rd Street, Philadelphia, PA 19104, USA}

\author{Tzu-Yung Huang}
\affiliation{Quantum Engineering Laboratory, Department of Electrical and Systems Engineering, University of Pennsylvania, 200 S. 33rd Street, Philadelphia, PA 19104, USA}

\author{Sander A. Mann}
\altaffiliation[Current address: ]{Department of Electrical and Computer Engineering, University of Texas at Austin, Austin, TX 78701, USA}
\affiliation{Center for Nanophotonics, AMOLF, Science Park 104, 1098 XG Amsterdam, The Netherlands}

\author{David A. Hopper}
\affiliation{Quantum Engineering Laboratory, Department of Electrical and Systems Engineering, University of Pennsylvania, 200 S. 33rd Street, Philadelphia, PA 19104, USA}
\affiliation{Department of Physics and Astronomy, University of Pennsylvania, 209 S. 33rd Street, Philadelphia, PA 19104, USA}

\author{Annemarie L. Exarhos}
\altaffiliation[Current address: ]{Department of Physics, Lawrence University, Appleton, WI 54911, USA}
\affiliation{Quantum Engineering Laboratory, Department of Electrical and Systems Engineering, University of Pennsylvania, 200 S. 33rd Street, Philadelphia, PA 19104, USA}

\author{Gerald G. Lopez}
\affiliation{Singh Center for Nanotechnology, University of Pennsylvania, 3205 Walnut St., Philadelphia, PA 19104, USA}

\author{Erik C. Garnett}
\affiliation{Center for Nanophotonics, AMOLF, Science Park 104, 1098 XG Amsterdam, The Netherlands}

\author{Lee C. Bassett}
\email[Corresponding author: ]{lbassett@seas.upenn.edu}
\affiliation{Quantum Engineering Laboratory, Department of Electrical and Systems Engineering, University of Pennsylvania, 200 S. 33rd Street, Philadelphia, PA 19104, USA}

\date{\today}


\def \PillarCalc {
\begin{figure}[!b]
\includegraphics[width=\textwidth]{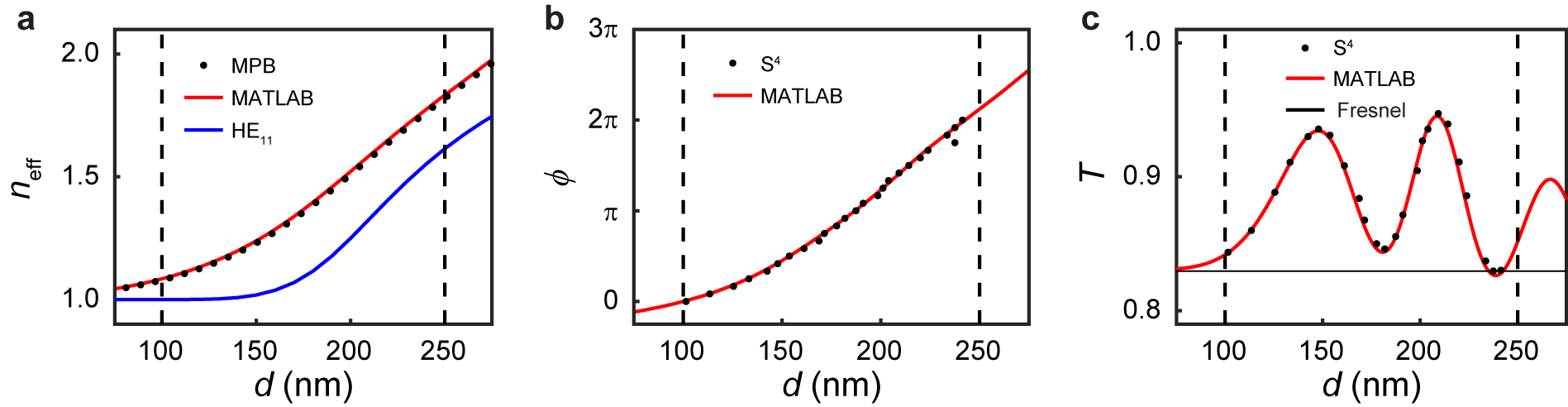}
\caption{\label{Fig:PillarCalc}\textbf{Metasurface element design | } Simulations of  (\textbf{a}) effective refractive index, $n_{\text{eff}}$, (\textbf{b}) optical transmission phase shift, $\phi$, and (\textbf{c}) transmission efficiency as a function of pillar diameter, $d$, for normal incidence at $\lambda = \SI{700}{\nano\meter}$ on a $\Lambda = \SI{300}{\nano\meter}$ grid. The effective index of an isolated waveguide is also shown in (\textbf{a}) and the theoretical transmission efficiency of a planar air/diamond at normal incidence is shown in (\textbf{c}). Black-dashed lines indicate the range of $d$ used for our metalens design. Comparisons of our MATLAB code, based on \cite{Rumpf_PIERS_11}, to open source simulation tools S$^4$ (see ref.~\cite{Liu_CPC_12}) and MIT Photonic Bands (MPB, see ref.~\cite{Johnson_OE_01}) are shown.}
\end{figure}
}

\def \MLExpSetup {
\begin{figure}[!t]
\includegraphics[width=\textwidth]{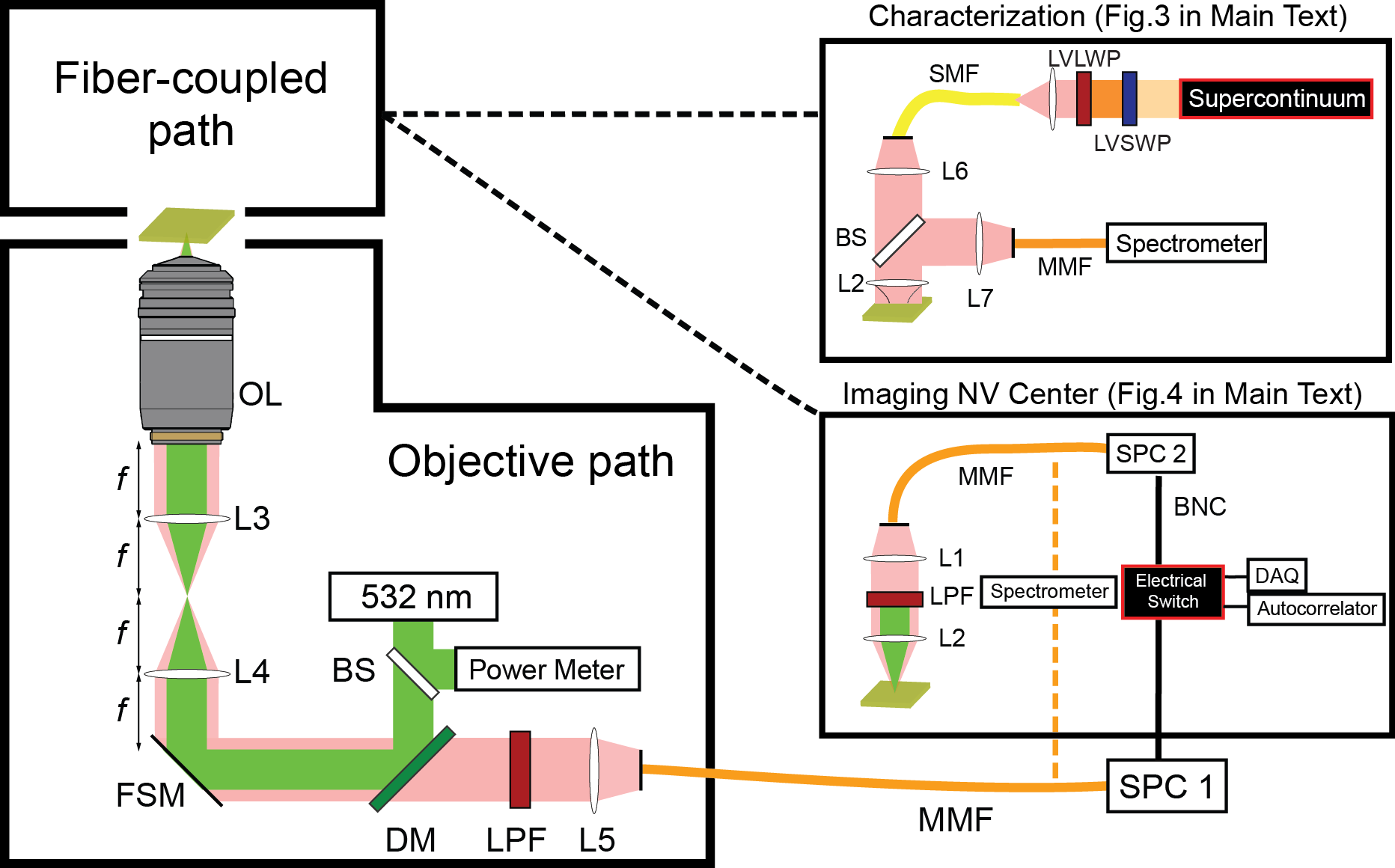}
\caption{\label{Fig:MLExpSetup} \textbf{Metalens Characterization and Imaging setup | } Experimental setup for characterization of the metalens and imaging an NV center through a confocal microscope. The setup is divided into the fiber-coupled path - where the metalens is coupled to either a single- or multi-mode fiber - and the objective path which enables confocal excitation and collection through an oil-immersion objective. The fiber-coupled path is modified to allow different experiment configurations with the metalens, as shown in the figure. The details of the full setup are discussed in the text.}
\end{figure}
}

\def \MLQELimage {
\begin{figure}[!h]
\includegraphics[width=\textwidth]{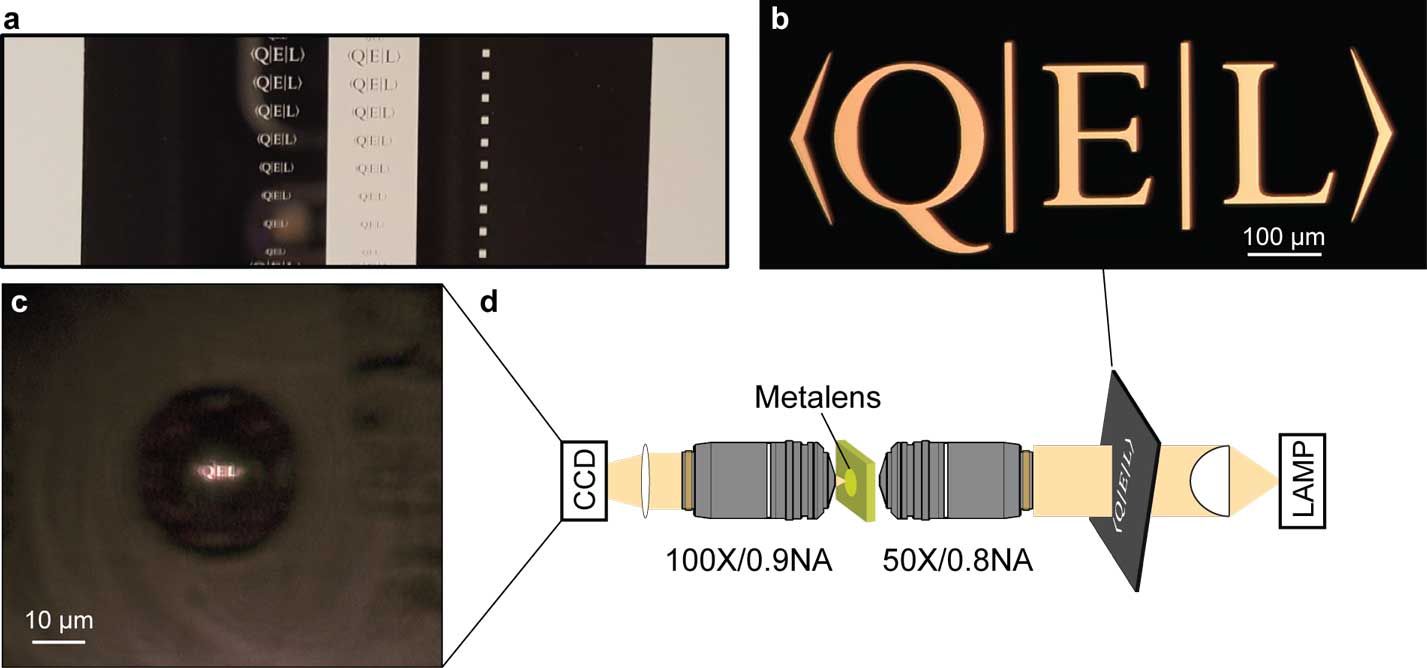}
\caption{\label{Fig:MLQELimage}\textbf{Imaging using the metalens | } \textbf{a}, Photograph of the chromium shadow mask fabricated on a 3"x1" glass slide. \textbf{b}, Bright-field microscope image of the shadow mask pattern imaged in (\textbf{c}) through the metalens using the setup in (\textbf{d}).}
\end{figure}
}

\def \FocusInAir {
\begin{figure}[!t]
\includegraphics[width=\textwidth]{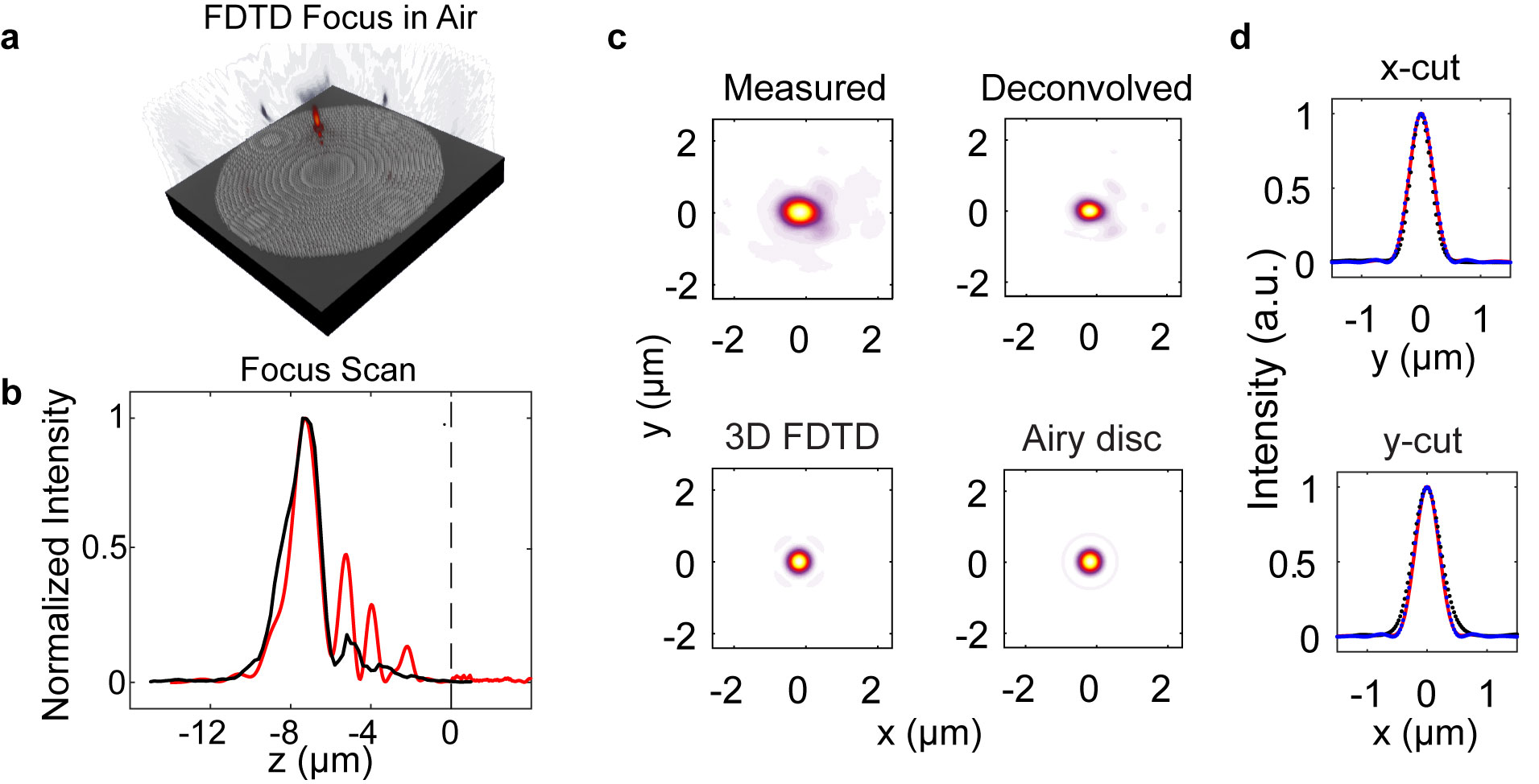}
\caption{\label{Fig:FocusInAir}\textbf{Simulated and Characterized Focus in Air |} \textbf{a}, FDTD simulation of the metalens focus in the air. \textbf{b}, FDTD (red) and measured (black) focus scan with the FSM at the center of the metalens. Dash line at $z = 0$ denotes surface of the metalens. The objective focus is moved further away from the metalens as $z$ becomes more negative. \textbf{c}, Measured focus spot plotted alongside deconvolved, FDTD, and airy disc focus spots. \textbf{d}, $x-$ and $y-$axis cross-section at focus of the deconvolved spot (black), the airy disc (blue), and FDTD simulation (red).}
\end{figure}
}

\def \MLwithNVBkgd {
\begin{figure}[!h]
\includegraphics[width=\textwidth]{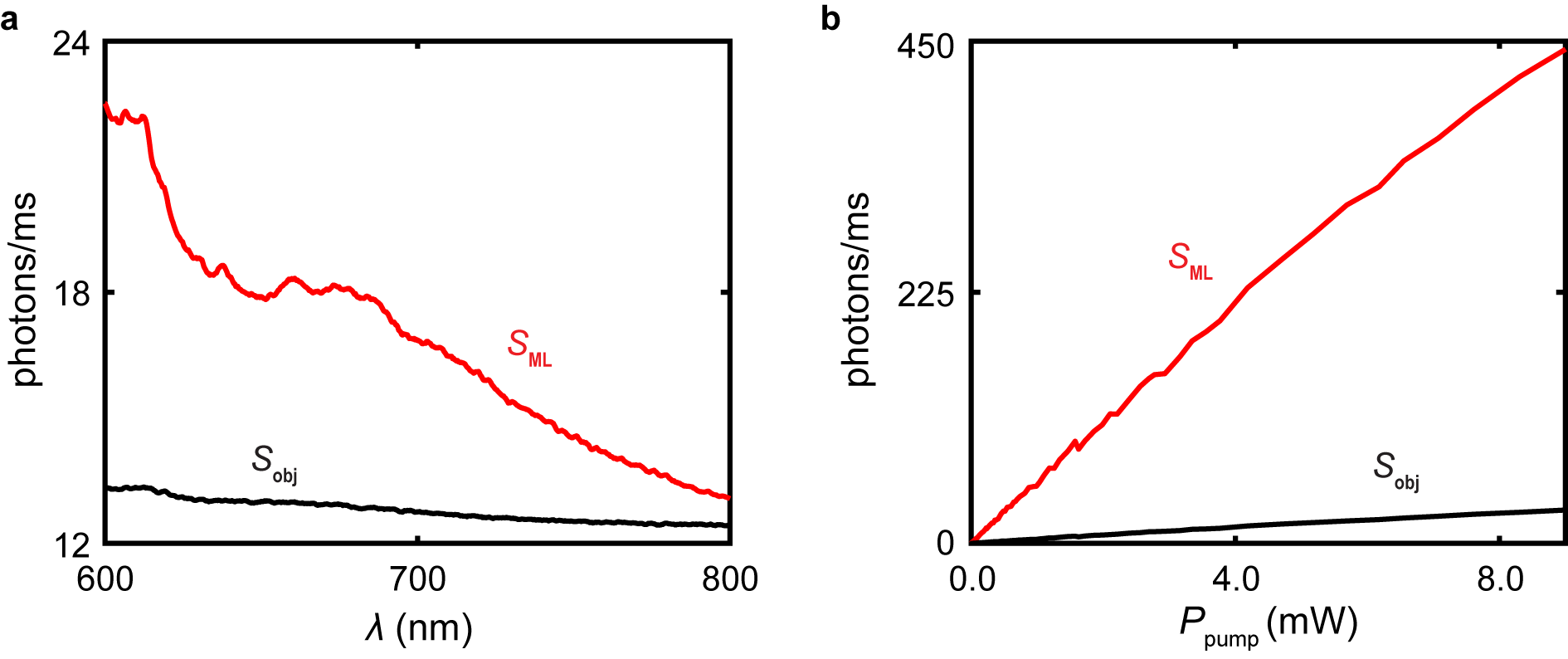}
\caption{\label{Fig:MLwithNVBkgd}\textbf{Background subtraction for PL measurements |} Background measurements for (\textbf{a}) PL spectra and (\textbf{b}) saturation curves. $S_{\text{ML}}$ and $S_{\text{obj}}$ denote signal from the metalens and objective paths, respectively.}
\end{figure}
}

\def \MLwithNVCorrBkgd {
\begin{figure}[!t]
\includegraphics[width=\textwidth]{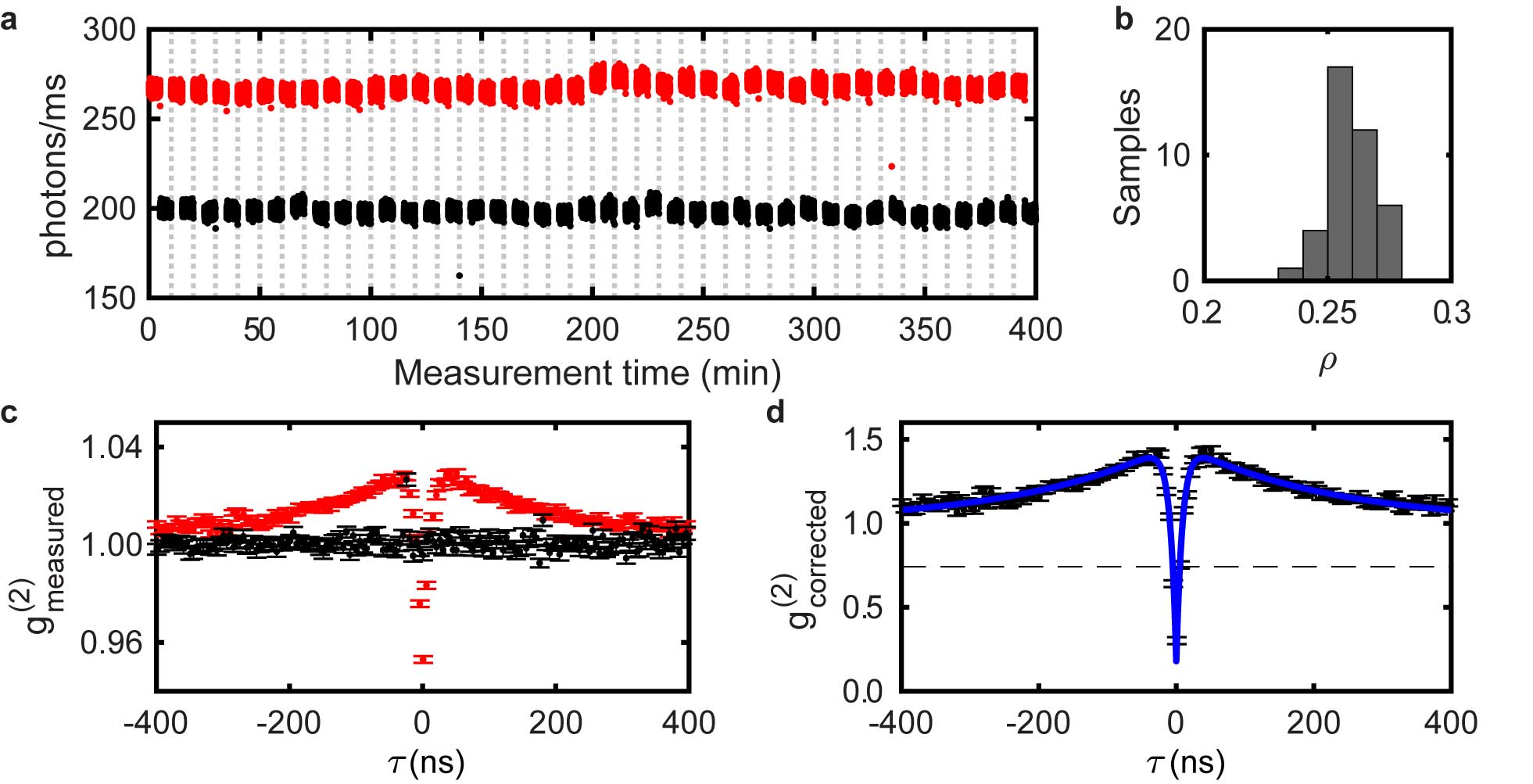}
\caption{\label{Fig:MLwithNVCorrBkgd}\textbf{Background subtraction and fitting for autocorrelation measurements |} \textbf{a}, Total countrates (photons/ms) as a function of overall measurement time, where each pair of experiments, separated by dashed lines, comprised of an on-NV measurement (red) and off-NV measurement (black). The countrates plotted here represent the sum of photons from both collection paths. \textbf{b}, Histogram of signal-to-background ratio, $\rho$, calculated from each pair of experiments described in \textbf{a}, using Eqn.~\ref{Eqn:rhoDef}. \textbf{c}, Cross-correlation of photons collected during on-NV measurements (red) and off-NV measurements (black). \textbf{d}, Background-corrected cross-correlation with antibunching below the single-emitter threshold at $\tau = 0$ and bunching characteristic of an NV center, fitted to a 3-level system correlation function.}
\end{figure}
}

\def \FSMCal {
\begin{figure}[!t]
\includegraphics[width=\textwidth]{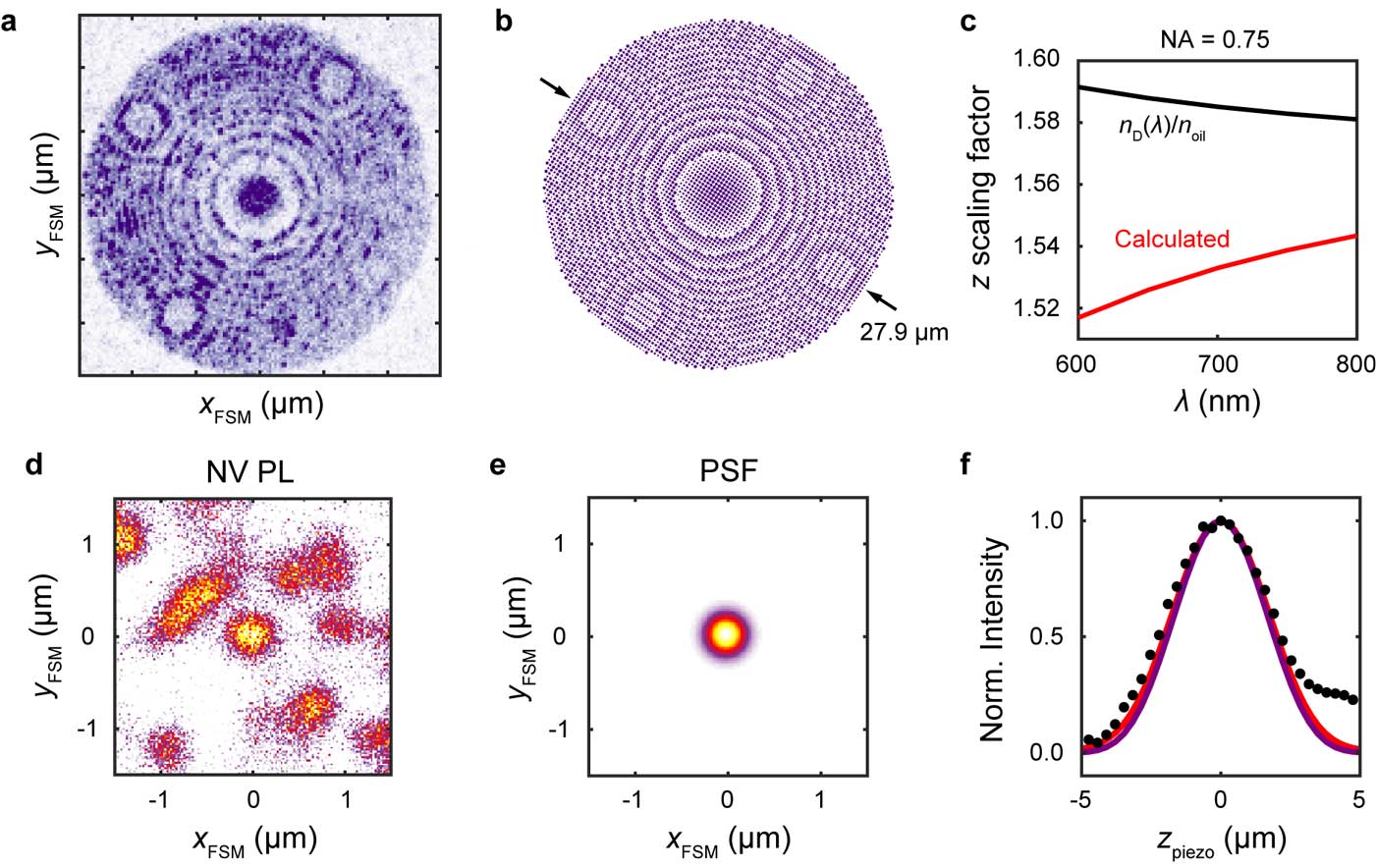}
\caption{\label{Fig:SetupCal}\textbf{Scanning microscope calibration |} \textbf{a}, PL scan of metalens surface, used to calibrate the fast-steering mirror (FSM). \textbf{b}, CAD layout of pillars. \textbf{c}, Objective focus position shift with $\hat{z}$-piezo stage movement.  \textbf{d}, PL scan through the confocal path with the $\hat{z}$-piezo stage positioned at the metalens focus. \textbf{e}, Fit to an isolated NV, used to determine NA$_{\text{obj}} \approx 0.75$.  \textbf{f}, Axial scan of an isolated NV (black circles), ideal axial response (purple curve) and calculated axial response with spherical aberration (red curve).}
\end{figure}
}

\def \TransverseFocusFits {
\begin{figure}[!t]
\includegraphics[width=\textwidth]{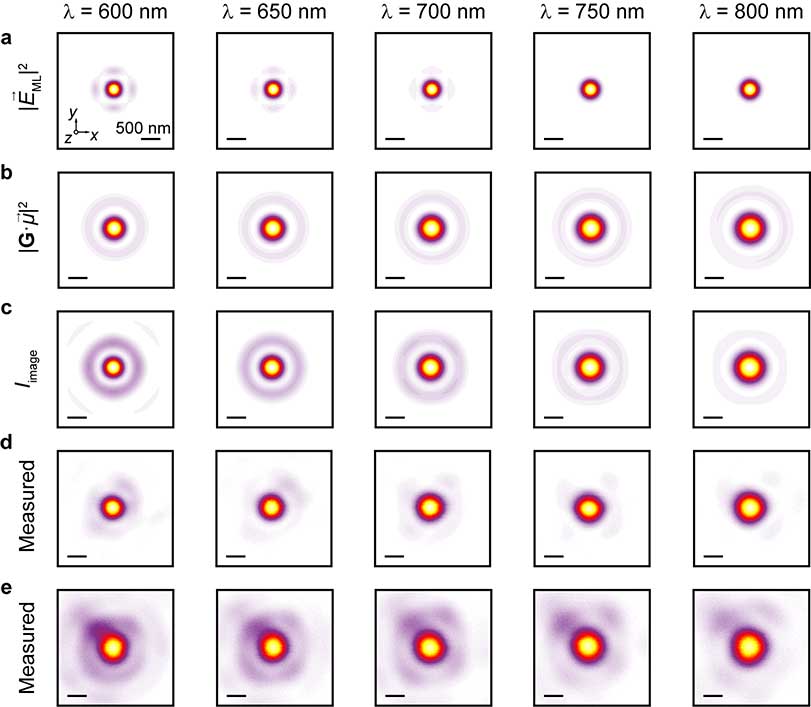}
\caption{\label{Fig:FocusFits}\textbf{Focus spot measurements |} \textbf{a}, Simulated focus at five wavelengths.  \textbf{b}, Microscope point spread function with spherical aberration caused by imaging through diamond included. \textbf{c}, Coherent convolution of simulated focus with microscope point-spread function. \textbf{d}, Measured focus spot. \textbf{e}, Focus spots measured with a longer focal length tube lens, effectively increasing the size of the confocal aperture.  All scale bars correspond to \SI{500}{\nano\meter}}
\end{figure}
}

\def \TransverseFocusCuts {
\begin{figure}[!t]
\includegraphics[width=\textwidth]{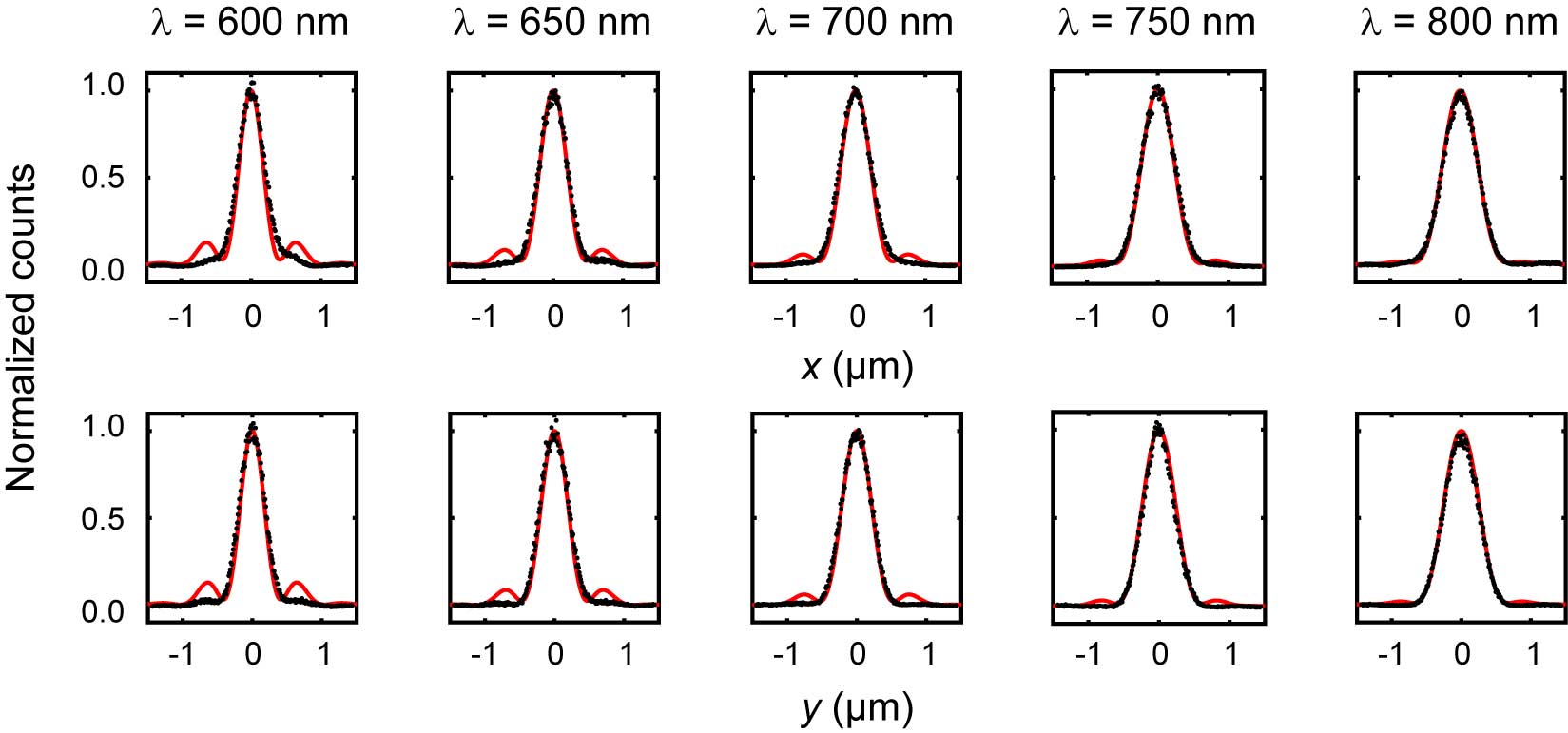}
\caption{\label{Fig:FocusCuts}\textbf{Focus fit cross-sections |} Transverse cross-sections of FDTD simulated focus spot convolved with the microscope PSF (red curves) and measured focus spot (black points).}
\end{figure}
}

\def \AxialFocusFits {
\begin{figure}[!t]
\includegraphics[width=\textwidth]{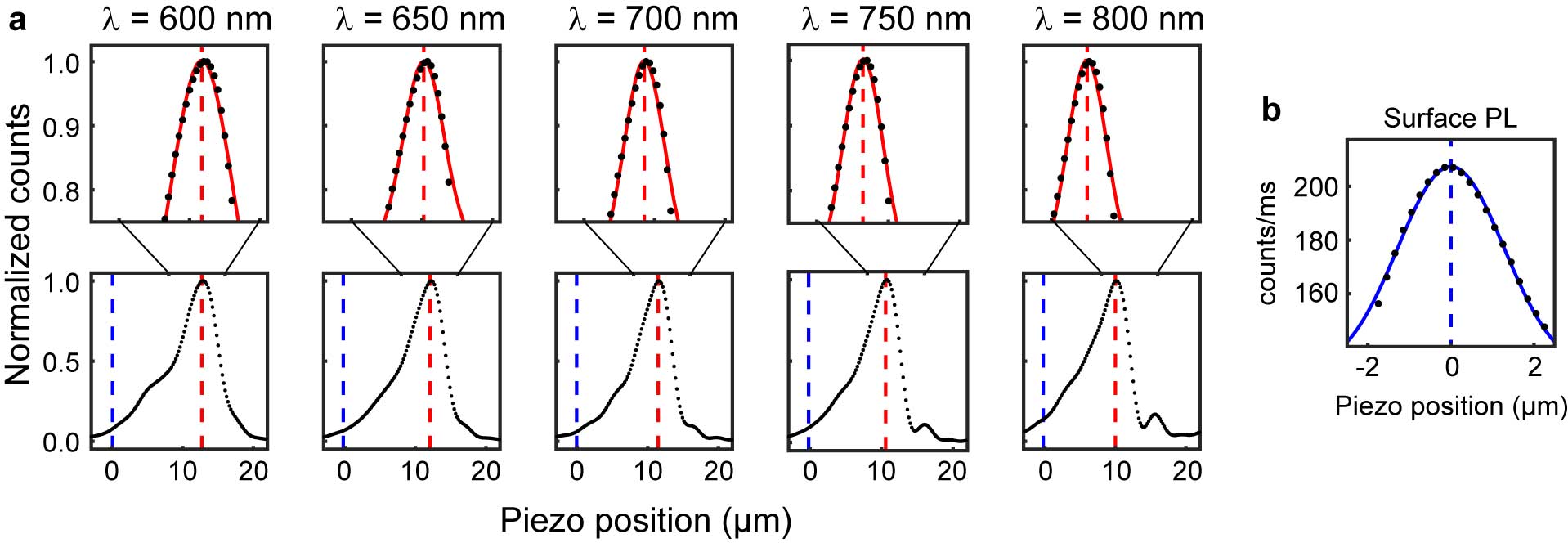}
\caption{\label{Fig:AxialScanFits}\textbf{Focal length measurements |} \textbf{a}, Normalized signal versus piezo stage position for five wavelengths.  \textbf{b}, Surface PL used to locate the sample surface.}
\end{figure}
}

\def \TowerPattern {
\begin{figure}[!t]
\includegraphics[width=\textwidth]{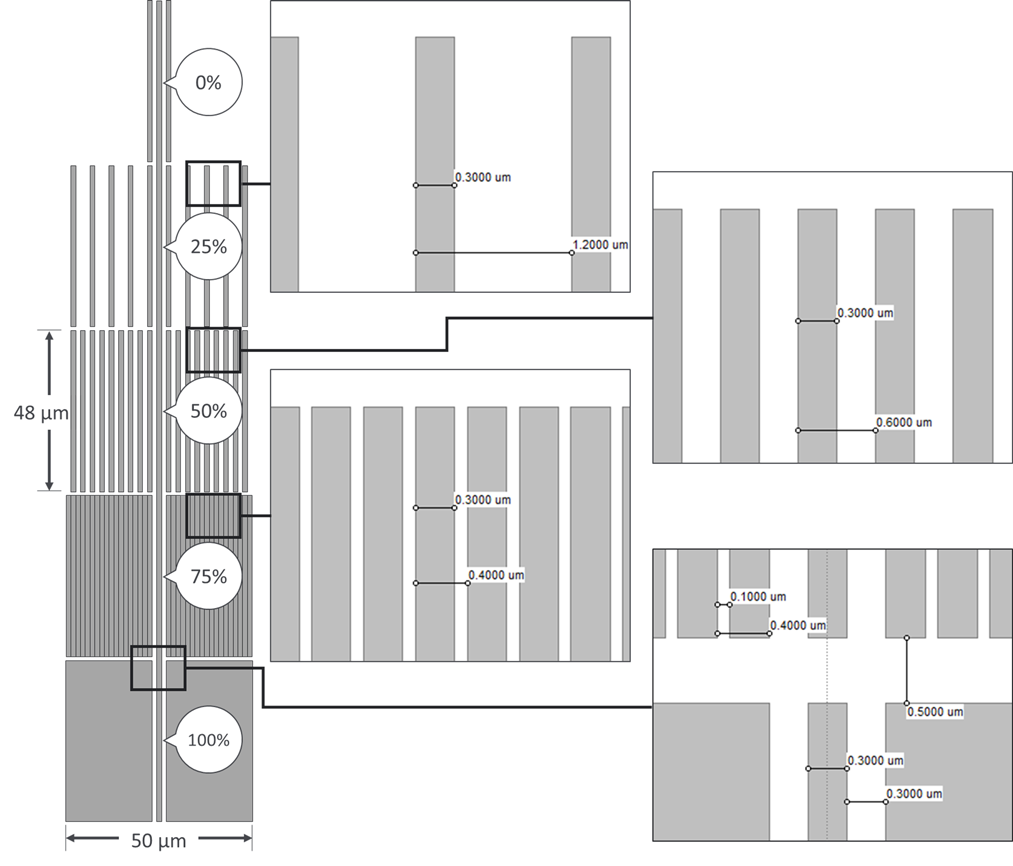}
\caption{\label{Fig:TowerPattern}\textbf{Tower Pattern | } A line-space tower pattern whose subsequent regions consist of 0\%, 25\%, 50\%, 75\% and 100\% pattern densities is illustrated. The pattern density, which is referenced at the center of the pattern, is determined by the pitch of each area and whose region measures over 4$\beta$ by 4$\beta$ in size. The corresponding line-width and pitch per pattern density is provided. The height of each region is \SI{48}{\micro\meter} tall and \SI{50}{\micro\meter} wide with a vertical gap of \SI{0.5}{\micro\meter} between regions. Measurements are performed in the center of each defined region.}
\end{figure}
}

\def \TowerPatternOM {
\begin{figure}[!t]
\includegraphics[width=\textwidth]{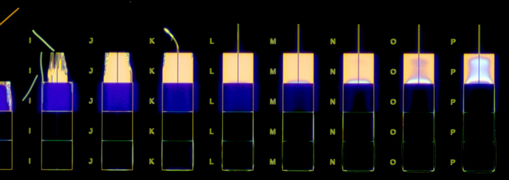}
\caption{\label{Fig:TowerPatternOM}\textbf{Exposure Latitude vs. Simulation | } Tower patterns written in HSQ atop Si seen using an optical microscope in a dark field mode. Exposure latitude data by post processing scanning electron microscope images of the line widths at different pattern densities across different doses.}
\end{figure}
}

\def \ExposureLatitude {
\begin{figure}[!t]
\includegraphics[width=\textwidth]{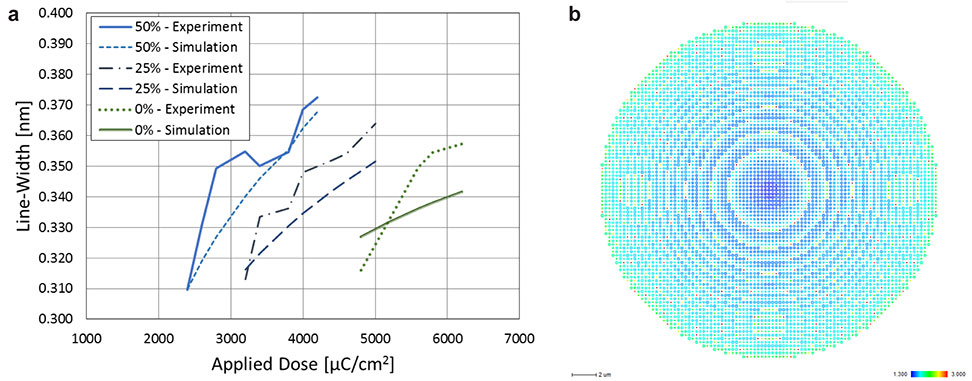}
\caption{\label{Fig:ExposureLatitude}\textbf{Exposure Latitude vs. Simulation | } \textbf{a}, The exposure latitude curves for 0\%, 25\% and 50\% pattern densities. Electron beam simulation fits the empirical data with an RMS = \SI{8}{\nano\meter}. \textbf{b}, Using the parameters from Tab.~\ref{Tab:PEC}, PEC was applied to the metalens design. The colors indicate various dose factors that are necessary to print the features to size in HSQ.}
\end{figure}
}

\section*{Supplementary Information}
\section{Design and simulations}
\subsection{Subwavelength element simulations}
\PillarCalc
Comparisons of our rigorous coupled-wave analysis (RCWA) MATLAB code to open source planewave expansion \cite{Johnson_OE_01} and RCWA software \cite{Liu_CPC_12} are shown in Fig.~\ref{Fig:PillarCalc} to verify the accuracy of our calculations. As described in the methods section of the main text, the Bloch-mode effective index calculated by solving for the eigenvalues of Maxwell's equations in a truncated planewave basis with implicit periodic boundary conditions is shown in Fig.~\ref{Fig:PillarCalc}\textbf{a}.  The effective index of the lowest-order HE$_{11}$ mode supported by an isolated pillar is also shown for comparison. The corresponding normal-incidence phase shift for \SI{1.0}{\micro\meter}-high pillars on a homogeneous diamond substrate, calculated by $\phi(d) = \angle t(d)$, is shown in Fig.~\ref{Fig:PillarCalc}\textbf{b}. 

Among the advantages of our metasurface design is its high transmission efficiency.  Since the effective index of each pillar lies naturally between the refractive index of air and that of diamond (Fig.~1\textbf{b} of the main text), the pillars are inherently anti-reflective with an average transmission efficiency of $88.6\%$, which is higher than the $83\%$ transmission efficiency predicted for an air/diamond interface by normal incidence Fresnel coefficients (Fig.~\ref{Fig:PillarCalc}\textbf{c}).  

\subsection{Image reconstruction}
\label{Sec:ImageRecon}
The image formed by our microscope can be described by an electric-field amplitude vector, $\vec{E}_{\text{image}}(\vec{r}_{\text{image}})$, that is a function of the FSM and $\hat{z}$-piezo stage positions described by $\vec{r}_{\text{image}} = x_{\text{FSM}}\cdot\hat{x} + y_{\text{FSM}}\cdot\hat{y} + z_{\text{piezo}}\cdot\hat{z}$.  This field vector can be expressed as a volume integral over the tensor Green's function (see ref. \cite{Novotny_12}), $\mathbf{G}(\vec{r}_{\text{image}},\vec{r})$, that describes the impulse response of our imaging system with a current distribution $\vec{J}(\vec{r})$ in the object space, $\vec{r}$, 

\begin{equation}
\vec{E}_{\text{image}}(\vec{r}_{\text{image}}) = j\omega\mu_0 \int_V \mathbf{G}(\vec{r}_{\text{image}},\vec{r})\cdot\vec{J}(\vec{r}) \text{d}V.
\label{Eqn:GreenField}
\end{equation}

\noindent Calculation of the tensor Green's function is described in the methods section.  

For the measurements described in Fig.~3 of the main text, the current distribution being imaged can be described by a displacement current, $\vec{J} = -j\omega\epsilon_{\text{D}}\epsilon_0\vec{E}$, caused by the focused fields of the metalens.  Substituting this into Eqn.~(\ref{Eqn:GreenField}) and normalizing to remove scaling factors, we find an expression for the image field:

\begin{align}
\vec{E}_{\text{image}}(\vec{r}_{\text{image}}) &= \nonumber \\
 \int_z \int_y \int_x &\mathbf{G}(x-x_{\text{FSM}},y-y_{\text{FSM}},z-z_{\text{piezo}})\cdot\vec{E}(x,y,z) \mathop \text{d}x \mathop \text{d}y \mathop \text{d}z \\
&= (G_{xx}*E_x + G_{xy}*E_y + G_{xz}*E_z)\cdot\hat{x} \nonumber \\
&+ (G_{yx}*E_x + G_{yy}*E_y + G_{yz}*E_z)\cdot\hat{y},
\label{Eqn:FieldFig3}
\end{align}

\noindent where $*$ denotes a three-dimensional spatial convolution.  The intensity calculated by the squared magnitude of Eqn.~(\ref{Eqn:FieldFig3}), $I = |\vec{E}_{\text{image}}|^2$ (Eqn.~(3) of the methods section) defines the image formed by our microscope for the measurements presented in Fig.~3 of the main text. 

For confocal PL measurements the current distribution in Eqn.~(\ref{Eqn:GreenField}) can be replaced by a dipole emitter excited by the \SI{532}{\nano\meter} pump laser, $\vec{J} = -j\omega \mathbf{\alpha}\cdot\vec{E}_{\text{pump}}(\vec{r}_{\text{image}},\vec{r},\lambda_{\text{pump}})\cdot\delta(\vec{r}=\vec{r}_0)$, where $\mathbf{\alpha}$ is the emitter polarizability tensor and $\delta(\vec{r}=\vec{r}_0)$ is the Dirac delta function representing a dipole located at $\vec{r}=\vec{r}_0$. Following the analysis of ref. \cite{Novotny_12}, the integrals in Eqn.~(\ref{Eqn:GreenField}) can be normalized and approximated as the incoherent product of the PSF at pump and PL wavelengths:

\begin{equation}
I \approx |I_0(\lambda_{\text{pump}})|^2\cdot|I_0((\lambda_{\text{PL}})|^2,
\label{Eqn:NVPSF}
\end{equation}

\noindent where $I_0$ is the lowest-order diffraction integral.  In the paraxial limit $I_0$ takes the form of an Airy disk in the transverse plane,

\begin{equation}
I_0 = \frac{2J_1\left(\text{NA}_{\text{obj}} k_0(r_{\text{image}}-r_0) \right)}{\text{NA}_{\text{obj}}  k_0(r_{\text{image}}-r_0)}
\label{Eqn:paraxialPSF}
\end{equation}

\noindent with $(r_{\text{image}}-r_0) = \sqrt{(x_{\text{FSM}}-x_0)^2 + (y_{\text{FSM}}-y_0)^2}$, $k_0 = 2\pi/\lambda$, and NA$_{\text{obj}}$ is the numerical aperture of the imaging objective.  Equations (\ref{Eqn:NVPSF}) and (\ref{Eqn:paraxialPSF}) are used to characterize the transverse response of our microscope in Sec.~\ref{Sec:Calibration}, while the axial response formed by scanning the piezo stage is described by evaluating Eqn.~(\ref{Eqn:NVPSF}) as a function of $z_{\text{piezo}}$:

\begin{equation}
I(x_{\text{FSM}} = 0,y_{\text{FSM}} = 0, z_{\text{piezo}}) = \left|\text{sinc}\left(\frac{\text{NA}_{\text{obj}}^2z_{\text{piezo}}}{2n_{\text{oil}}^2\lambda} \right)\right|^4,
\label{Eqn:AxialRes}
\end{equation}

\noindent where $n_{\text{oil}} = 1.518$ is the refractive index of the immersion oil used with our objective.



\section{Electron Beam Lithography Methods: Process Characterization, Data Preparation and Proximity Effect Correction}
An Elionix ELS-7500EX 50 keV electron beam lithography (EBL) tool was used to generate the metalens pattern in hydrogen silsesquioxane (HSQ), a common negative tone EBL resist, atop diamond. Using a \SI{300}{\micro\meter} field size and a beam current of \SI{1}{\nano\ampere} on a \SI{5}{\nano\meter} beam step size (shot pitch), the final pattern was exposed as a direct result of careful process characterization and modeling. In this section, we will describe the patterns and methods to generate the  the proximity effect correction (PEC) parameters for  the metalens.

\TowerPattern

\subsection{Calibration Pattern}
To calibrate the resist process, a tower pattern of lines and spaces was exposed in a dose matrix. Illustrated in Fig.~\ref{Fig:TowerPattern} is the line and space tower pattern of various pitch representing 0\%, 25\%, 50\%, 75\% and 100\% pattern densities. According to Monte Carlo simulations performed using TRACER\cite{TRACER} by GenISys, exposing with a 50 keV tool atop Si yields a backscatter length ($\beta$) of \SI{10}{\micro\meter}. Therefore, each pattern density region is 4$\beta$ by 4$\beta$ or greater in size such that the center of the pattern, when exposed, has a total absorbed energy that is saturated from backscattered electrons. 

\TowerPatternOM

A specific pattern density is achieved by applying a specific pitch to the line-space pattern. For example, a 25\% pattern density consists of \SI{300}{\nano\meter} lines on a \SI{1200}{\nano\meter} pitch, where the line occupies 25\% of the full pitch. The line-width and pitch dimensions are provided in Fig.~\ref{Fig:TowerPattern}. After exposure and development, the final pattern seen in Fig.~\ref{Fig:TowerPatternOM} is imaged using a scanning electron microscope (SEM). The SEM images are post processed to extract the pattern density dependent exposure latitudes. 

\subsection{Process modeling and Correction}
PEC is an edge-correction technology in which the absorbed energy of the resist in the pattern is analyzed and dose assignments are made such that the absorbed energy at threshold lands at the edge of the intended design. This threshold is associated with the resist sensitivity and development chemistry. Densely written patterns build up additional absorbed energy \textit{via} electron backscatter, requiring a local dose reduction; conversely, sparsely written (low density) patterns require an increase in local dose. The amount of background energy at these dense and sparse pattern densities directly impacts the exposure latitude, which is the critical dimension response to a change in dose.

\ExposureLatitude

HSQ has been shown to exhibit non-ideal behavior in its response to proximity effect correction methods due to  microloading effects during resist development \cite{Bickford_JVSTB_14}. Using BEAMER\cite{BEAMER} by GeniSys, a genetic algorithm was employed to model the empirical exposure latitude data. For this simulation, only the 0\%, 25\% and 50\% pattern density data were of interest since the metalens pattern density falls within this range. Reducing the input data reduces the convergence time. The parameters used to obtain the model fit were the effective process blur, development bias, and base dose. These values are determined \textit{via} simulation in the genetic algorithm by matching the simulated resist edge contours to the experimental exposure latitude data obtained from the tower pattern in Fig.~\ref{Fig:TowerPattern} that was exposed in a dose matrix as shown in Fig.~\ref{Fig:TowerPatternOM}. The resulting effective process blur is then convolved into the electron point spread function to perform the simulation. The slope of the experimental exposure latitude data is matched in simulation by changing the effective process blur accordingly (Fig.\ref{Fig:ExposureLatitude}\textbf{a}). By adding two extra degrees of freedom, development bias and base dose, the algorithm can converge properly.

\begin{table}[h]
\centering
\begin{tabular}{|c|c|}
\hline
$\alpha$ & \SI{5}{\nano\meter} \\
\hline
$\beta$ & \SI{10}{\micro\meter} \\
\hline
Effective Blur & \SI{67}{\nano\meter} \\
\hline
Bias & -\SI{5}{\nano\meter}\\
\hline
\end{tabular}
\caption{Proximity Effect Correction Parameters\label{Tab:PEC}}
\end{table}

The final pattern was proximity effect corrected using the parameter found in Tab.~\ref{Tab:PEC}. As a result, the metalens is fractured such that the shapes receive the appropriate dose to print the features to size (Fig.~\ref{Fig:ExposureLatitude}\textbf{b}).

\section{Imaging with the metalens}
\MLQELimage
The HPHT diamond hosting the metalens is placed in a conventional upright microscope (Olympus, BX41) for brightfield transmission and reflection imaging (Fig.~2 of the main text). The bright-field transmission microscope image in Fig.~2\textbf{d} of the main text was created by placing a chromium shadow mask between a lamp and a focusing objective, which was focused through the metalens and imaged on a CCD using a second objective as described in Fig.~\ref{Fig:MLQELimage}.  The shadow mask was fabricated by e-beam depositing chromium on a glass microscope slide (Fig.~\ref{Fig:MLQELimage}\textbf{a}), and creating the pattern shown in Fig.~\ref{Fig:MLQELimage}\textbf{b} with a combination of photolithography and chemical etching.  The resulting CCD image shown in Fig.~\ref{Fig:MLQELimage}\textbf{c} was created using the transmission microscope shown in Fig.~\ref{Fig:MLQELimage}\textbf{d}.  

\section{Metalens characterization}

\subsection{Measurement setup}
\MLExpSetup
The diamond is mounted on a glass cover slip, which is attached to the stage of a custom-built laser-scanning confocal microscope (Fig.~\ref{Fig:MLExpSetup}) for characterization and NV center imaging (Figs.~3,4 of the main text). The laser-scanning confocal microscope has two optical paths for simultaneously probing the metalens from air and through the diamond substrate: a fiber-coupled path and an objective path. The objective path consists of a $4f$ relay-lens system with achromatic doublet lenses (L3 and L4, Newport, $\SI{25.4}{\milli\meter}\times\SI{150}{\milli\meter}$ focal length, PAC058AR.14), which is used to align the back aperture of the objective to a fast-steering mirror (FSM, Optics in motion, OIM101). This is followed by a \SI{560}{\nano\meter} long-pass dichroic mirror (Semrock, BrightLine FF560-FDi01) which directs the \SI{532}{\nano\meter} excitation laser (Coherent, Compass 315M-150) into the objective (OL, Nikon, Plan Flour x100/0.5-1.3) while wavelengths above \SI{560}{\nano\meter} are passed through a \SI{532}{\nano\meter} and a \SI{568}{\nano\meter} long-pass filter (Semrock, EdgeBasic BLP01-532R, EdgeBasic BLP01-568R) before being focused down to a \SI{25}{\micro\meter}-core, 0.1~NA, multimode fiber (Thorlabs M67L01) \textit{via} the achromatic doublet lens (L5, Newport, $\SI{25.4}{\milli\meter}\times\SI{50}{\milli\meter}$ focal length, PAC049AR.14). The multimode fiber is then connected to a single-photon counting module, (SPCM, Excelitas, SPCM-AQRH-14-FC) or a spectrometer (Princeton Instruments, IsoPlane-160, \SI{750}{\nano\meter} blaze wavelength with 1200 G/mm) with a thermoelectrically-cooled CCD (Princeton Instruments PIXIS 100BX). The electrical output of the single-photon counting module is routed via BNC cables to either a data acquisition card (DAQ, National Instruments PCIe-6323) or a time-correlated single-photon counting card (PicoQuant, PicoHarp 300).

The fiber-coupled path is modified to enable different experiments conducted on the metalens. For characterization, a broadband supercontinuum source (Fianium WhiteLase SC400) was coupled into a single-mode fiber (Thorlabs P1-630AR-2). A $f = \SI{2.0}{\milli\meter}$ collimating lens (L6, Thorlabs CFC-2X-A) was used to create a \SI{380}{\micro\meter} diameter Gaussian beam that emulates the planewave source used in our FDTD simulations. The excitation wavelength is set by passing the supercontinuum beam through a set of linear variable short-pass (Delta Optical Thin Film, LF102474) and long-pass filters (Delta Optical Thin Film LF102475) prior to fiber-coupling, which can be adjusted to filter out a single wavelength with $<\SI{8}{\nano\meter}$ bandwidth or be removed completely for broadband excitation. For reflectance measurements, a $f = \SI{15}{\milli\meter}$ achromatic doublet lens (L2, Thorlabs, AC064-015-B) is added to focus the collimated excitation beam to a $\sim\SI{30}{\micro\meter}$-diameter spot at the top surface of the diamond.  A beamspliter cube (Thorlabs, BS014) was added between the collimating and focusing lenses so that reflected light could be focused into a \SI{200}{\micro\meter}-core MMF (Thorlabs, M25L01) that is coupled to a spectrometer (Thorlabs CCS100) using a $f = \SI{100}{\milli\meter}$ achromatic doublet lens (L7, Newport, PAC052AR.14). To modify this setup for imaging an NV center, a \SI{532}{\nano\meter} and a \SI{568}{\nano\meter} long-pass filter (Semrock, EdgeBasic BLP01-532R, EdgeBasic BLP01-568R) is placed after L2 and the filtered light is focused down to a \SI{25}{\micro\meter}-core, 0.1~NA, multimode fiber (Thorlabs M67L02) with a $f = \SI{13}{\milli\meter}$ achromatic doublet lens (L1, Thorlabs, AC064-013-B). The multimode fiber can then be connected to a single-photon counting module or a spectrometer as described in the previous paragraph.

\subsection{Calibration}
\label{Sec:Calibration}
Calibration of the fast-steering mirror (FSM) is critical for characterization of the metalens's point-spread function at focus. To perform this calibration, a PL scan of the metalens surface was taken with \SI{532}{\nano\meter} pump beam (Fig.~\ref{Fig:SetupCal}\textbf{a}), and the image was compared to the CAD layout of the metalens pattern (Fig.~\ref{Fig:SetupCal}\textbf{b}) to determine the differential voltage required to move the FSM by a known distance in $x$ and $y$.
\FSMCal

The relative shift in axial position of the confocal collection volume caused by piezo stage movements is scaled by a factor ranging from $\frac{n_{\text{D}}}{n_{\text{oil}}}$ to $\frac{n_{\text{D}} \cos\theta_{\text{D}}}{n_{\text{oil}}\cos\theta_{\text{oil}}}$, where $\theta_{\text{D,oil}} = \sin^{-1}\left(\frac{\text{NA}}{n_{\text{oil,diamond}}}\right)$ are the maximum focusing angles in diamond and oil, respectively\cite{Visser_Scanning_94}.  We calculate this scaling factor using our numerical PSF model described in the methods section, and find that it is $\approx \frac{n_{\text{D}}}{n_{\text{oil}}}$ (Fig.~\ref{Fig:SetupCal}\textbf{c}), which is applied to the measured piezo stage position, $z'_{\text{piezo}}$, to find the physical displacement of the confocal volume within the sample, $z_{\text{piezo}} \approx \frac{n_{\text{D}}}{n_{\text{oil}}} z'_{\text{piezo}}$. The dispersive refractive index of diamond, $n_{\text{D}} (\lambda)$, used for the calculations in Fig.~\ref{Fig:SetupCal}\textbf{c} was modeled using the Sellmeier equation with coefficients from ref. \cite{Mildren_OED_Ch1_13}.  The sample thickness was checked by focusing $\SI{532}{\nano\meter}$ on both the bottom surface and top surface of the diamond, and measuring the relative position on the piezo stage.  The piezo stage displacement was $\SI{92}{\micro\meter}$, and the iris of the objective was set to NA$_{\text{obj}} = 0.5$.  The numerically calculated scaling factor 1.6, giving a sample thickness of \SI{147}{\micro\meter}. 

The objective lens used has an adjustable iris, which effectively reduces the NA to improve spherical aberration.  The collar was set to NA$_{\text{obj}}\approx 0.75$, which was confirmed by measuring PL from an NV center (Fig.~\ref{Fig:SetupCal}\textbf{d}), and fitting the PL scan as an incoherent convolution of two Airy disks using Eqns.~(\ref{Eqn:NVPSF},\ref{Eqn:paraxialPSF}). We found that the PSF of our microscope was not limited by the spot size of the excitation beam (either due to operating at saturation, or the pump beam not being diffraction limited), and thus modified Eqn.~(\ref{Eqn:NVPSF}) to fit $I = |I_0(\lambda_{\text{PL}})|^2 \cdot |I_0(\lambda_{\text{PL}})|^2$, with the results shown in Fig.~\ref{Fig:SetupCal}\textbf{e}.  Fits were performed using both $\lambda_{\text{PL}} = \SI{700}{\nano\meter}$ and a weighted fit over the NV PL spectrum, resulting in fit values of NA$_{\text{obj}} = 0.76 \pm 0.03$ and NA$_{\text{obj}} = 0.73 \pm 0.03$, respectively.  Using the fit value of NA$_{\text{obj}}\approx 0.75$, the unaberrated axial PSF corresponding to Eqn.~(\ref{Eqn:AxialRes}) (purple curve) and numerically evaluated aberrated axial PSF (red curve) are compared to measurements (black circles) in Fig.~\ref{Fig:SetupCal}\textbf{f}.  



\subsection{Focal length}

\AxialFocusFits 

The focal length of the metalens shown in Fig.~3\textbf{e} of the main text (right axis) was measured at five wavelengths by setting the FSM position to the peak of the transverse focused spot and scanning the piezo sample stage in the $\hat{z}$ (axial)-direction by \SI{200}{\nano\meter} steps, corresponding to shifts of $\frac{n_{\text{D}}}{n_{\text{oil}}}\cdot\SI{200}{\nano\meter} \approx \SI{315}{\nano\meter}$ inside the diamond. The position of focus was determined by fitting the peak signal of the piezo scan at each wavelength to a Gaussian (Fig.~\ref{Fig:AxialScanFits}\textbf{a}). The bright PL of the metalens surface was also fit with a Gaussian (Fig.~\ref{Fig:AxialScanFits}\textbf{b}) and used to calibrate the relative distance between the sample surface and the metalens focus.  

The chromatic aberration of our imaging system was checked by feeding the supercontinuum through the collection line and measuring the location of the metalens's surface \textit{via} a CCD camera in the collection path. Since the supercontinuum is coupled to a SMF, we can achieve this by simply coupling the SMF to the MMF in the objective collection path with an FC-to-FC fiber connector (Thorlabs, ADAFC1). By verifying that the surface location is the same when the excitation source is band-passed to \SI{600}{\nano\meter} as when it is \SI{800}{\nano\meter}, the chromatic aberration of the system was found to be negligible.  

\subsection{Field profiles}

A comparison of the simulated metalens focus, $|\vec{E}_{\text{ML}}|^2$, microscope PSF, $|\mathbf{G}\cdot\vec{p}|^2$, image formed by convolving the focus and PSF, $I_{\text{image}}$, and two sets of measured data at five wavelengths from \SIrange{600}{800}{\nano\meter} are shown in Fig.~\ref{Fig:FocusFits}.  The microscope PSF is represented by the product of three dipole moments oriented along the three Cartesian axes, $\vec{p} = (\hat{x} + \hat{y} + \hat{z})\cdot\delta(\vec{r}_{\text{image}})$, and the tensor Green's function, $\mathbf{G}$, described in the methods section of the main text. The image intensity, $I_{\text{image}}$, has been calculated by a coherent convolution as described in Sec.~\ref{Sec:ImageRecon}.  The measurements were taken with two different tube lenses (L5 in Fig.~\ref{Fig:MLExpSetup}), $f = \SI{50}{\milli\meter}$ with a $6\times$ reducing telescope (Fig.~\ref{Fig:FocusFits}\textbf{d}), and $f = \SI{100}{\milli\meter}$ (Fig.~\ref{Fig:FocusFits}\textbf{e}).  Changing the tube lens effectively changes the size of the collection aperture relative to the image size.  For the measurements taken in Fig.~\ref{Fig:FocusFits}\textbf{d}, the aperture could be considered infinitesimal (i. e., far below the confocal condition) and does not affect the imaging resolution\cite{Corle_96}, whereas in Fig.~\ref{Fig:FocusFits}\textbf{e} the aperture is finite and decreases the resolution with which the spot is measured.


\TransverseFocusFits

Comparisons of $x$ and $y$ cross-sections of the convolved simulations (Fig.~\ref{Fig:FocusFits}\textbf{c}) and measurements with infinitesimal pinhole (Fig.~\ref{Fig:FocusFits}\textbf{d}) are shown in Fig.~\ref{Fig:FocusCuts}.  The agreement between model and measurement seen in Fig.~\ref{Fig:FocusCuts} is remarkable, given that there are no free parameters.  In other words, the plots in Fig.~\ref{Fig:FocusCuts} represent an agreement between theory and experiment, rather than a fit to experimental data.

\TransverseFocusCuts 




\subsection{Focusing in air}

To measure the focus spot formed in air when the metalens is illuminated by a collimated beam from inside of the diamond (Fig.~\ref{Fig:FocusInAir}\textbf{a}), the diamond substrate is mounted upside-down on the inverted microscope shown in Fig.~\ref{Fig:MLExpSetup} with the metalens facing downwards towards a 100x air objective (Olympus, UMPlanFl $100\times$/0.90) in the objective path. A \SI{633}{\nano\meter} He-Ne laser source (Melles Griot 05-LHP-153) is SMF-coupled and collimated \textit{via} a $f = \SI{2.0}{\milli\meter}$ collimating lens (Thorlabs CFC-2X-A) to illuminate the back-side of the diamond substrate from the fiber-coupled path. The methods for measuring the focus spots and focal length of the metalens are described in the Experimental section of Methods, as the collection path after the objective is identical to the objective path shown in Fig.~\ref{Fig:MLExpSetup}. An axial scan of the metalens focus in air is plotted in Fig.~\ref{Fig:FocusInAir}\textbf{b}, showing an excellent agreement with the FDTD simulation. The measured transverse focus spot, shown in Fig.~\ref{Fig:FocusInAir}\textbf{c}, is deconvolved using blind deconvolution with MATLAB's \textbf{deconvblind} command. The cross-sections of the deconvolved focus spot, plotted in Fig.~\ref{Fig:FocusInAir}\textbf{d} in black, demonstrate again an excellent agreement with the FDTD simulation (red). 
\FocusInAir

\section{Background subtraction for NV measurements}

\subsection{Spectra and Saturation curves}

Experimental setup for spectra and saturation curves measurements are described in Fig.~\ref{Fig:MLExpSetup}. Background spectra and saturation curves are measured at a transverse scan position that is away from the NV center but still within the field-of-view of the metalens. Signal (on-NV) and background (off-NV) spectra for both metalens and objective paths were collected with a 5-min acquisition time. The background spectra for both paths, plotted in Fig.~\ref{Fig:MLwithNVBkgd}\textbf{a}, are subtracted from the signal spectra to yield the points plotted in Fig.~4\textbf{d} in the main text. For saturation curves, the \SI{532}{\nano\meter} pump beam is passed through a variable optical-density filter (Thorlabs, NDC-50C-4), before going through a beamsplitter cube (Thorlabs BS014) which enables the pump beam's power to be measured by a power meter (Thorlabs PM100D). For each power increment, signal and background photon counts were measured for \SI{500}{\milli\second} for both the metalens and objective paths. The background countrates, plotted in Fig.~\ref{Fig:MLwithNVBkgd}\textbf{b}, are subtracted from the signal countrates to yield the points plotted in Fig.~4\textbf{e} in the main text. 

\MLwithNVBkgd

\subsection{Autocorrelation}
When we centered the FSM on the NV center to record photons for cross-correlation, we are collecting both the photons emitted from the NV center as well as photons from the background. To account for this background and correct for it, we need to examine the $g^{(2)}(\tau)$ function and its boundary conditions. Given an arbitrary correlation function mixed with Poissanian background, the measured correlated function, $g^{(2)}_{\text{measured}}(\tau)$, is related to the ideal, background-free correlated function, $g^{(2)}_{\text{ideal}}(\tau)$, in the following way:

\begin{equation}
g^{(2)}_{\text{measured}}(\tau) = 1 - \rho^2 + \rho^2g^{(2)}_{\text{ideal}}(\tau) 
\label{Eqn:gMeasured}
\end{equation}

\noindent this adjusted the boundary conditions of $g^{(2)}_{\text{measured}}(\tau)$ to the following:

\begin{equation}
g^{(2)}_{\text{measured}} = 
\begin{dcases}
 1 - \rho^2 & \tau = 0\\
1 & \tau = \infty 
\end{dcases}
\label{Eqn:corrMeasBound}
\end{equation}

\noindent where $\rho$ is defined as:

\begin{equation}
\rho = \dfrac{S}{S + B} = 1 - \dfrac{B}{S + B}
\label{Eqn:rhoDef}
\end{equation}

\noindent where $S$ is the signal and $B$ is the background. Both Eqn.~\ref{Eqn:gMeasured} and Eqn.~\ref{Eqn:rhoDef} make the assumption that the background in the measurement is Poissanian. To justify this assumption, we moved the FSM to a spot off the NV center that is still within the metalens' field of view and measured photons from both metalens and objective paths for the same duration as we did for when the FSM is centered on the NV center (5 minutes). The off-NV (background) measurement was performed immediately following the on-NV (signal) measurement and the pair of measurements was repeated for 40 times. The recorded countrates are shown in Fig.~\ref{Fig:MLwithNVCorrBkgd}\textbf{a} illustrating the consistency and stability of countrates over more than six hours of measurements.

We calculate $\rho$ for each pair of experiments by using Eqn.~\ref{Eqn:rhoDef} where $B$ is measured as countrates from off-NV measurements and $S + B$ is measured as countrates from on-NV measurements. The distribution of these $\rho$ values is plotted in Fig.~\ref{Fig:MLwithNVCorrBkgd}\textbf{b}. 

Next, we calculate the cross-correlation of the recorded photons in the signal as well as the background measurements, shown in Fig.~\ref{Fig:MLwithNVCorrBkgd}\textbf{c}.
We use a variant of the algorithm developed by Laurence \textit{et al.} \cite{Laurence2006}, to calculate the cross correlation function from the raw photon arrival times.  These measurements clearly demonstrate that the background is Poissonian, whereas the background-incorporated signal measurements showed cross-correlation characteristic of a single- or few-photon emitter.
\MLwithNVCorrBkgd
To perform the background correction for $g^{(2)}_{\text{measured}}(\tau)$, we rearrange Eqn.~(\ref{Eqn:gMeasured}) to:

\begin{equation}
g^{(2)}_{\text{background-corrected}}(\tau) 
= \dfrac{g^{(2)}_{\text{measured}}(\tau) - (1 - \rho^2)}{\rho^2}
\label{Eqn:gCorrected}
\end{equation}

\noindent which yields the points plotted in Fig.~\ref{Fig:MLwithNVCorrBkgd}\textbf{d} and in the main text. 

We fit the background-corrected autocorrelation function using the well-known approximation of the NV center as a 3-level system\cite{Brouri_OL_00}: 

\begin{align}
g^{(2)}_{\text{background-corrected}}(\tau) &= 1 - Ae^{-\tfrac{\abs{t-t_0}}{\tau_1}} + Ce^{-\tfrac{\abs{t-t_0}}{\tau_2}}
\label{Eqn:corrFunc}
\end{align}

\noindent where ideally $A = C + 1$ but we allow for the possibility of $A < C + 1$ to account for imperfect background measurements and finite detector bandwidth. The results of this fit is plotted in Fig.~\ref{Fig:MLwithNVCorrBkgd}\textbf{d}, clearly showing the antibunching dip at $\tau = 0$ below $\dfrac{1 + C}{2}$ to satisfy the condition of a single-photon emitter, as well as the characteristic short-delay bunching of an NV-center due to shelving in the spin-singlet manifold.

\bibliography{Metalens_NP_17}